\documentclass{aa}
\usepackage[varg]{txfonts}
\usepackage{graphicx}
\usepackage{natbib,twoopt}
\usepackage[]{hyperref}

\begin{document}
   \title{Flying far and fast: the distribution of distant hypervelocity star 
          candidates from \textit{Gaia} DR2 data}
   \author{R. de la Fuente Marcos$^{1}$
           \and
           C. de la Fuente Marcos$^{2}$}
   \authorrunning{R. de la Fuente Marcos \and C. de la Fuente Marcos}
   \titlerunning{Distant hypervelocity star candidates from \textit{Gaia} DR2}
   \offprints{R. de la Fuente Marcos, \email{rauldelafuentemarcos@ucm.es}}
   \institute{$^1$AEGORA Research Group,
              Facultad de Ciencias Matem\'aticas,
              Universidad Complutense de Madrid,
              Ciudad Universitaria, E-28040 Madrid, Spain \\
              $^2$ Universidad Complutense de Madrid,
              Ciudad Universitaria, E-28040 Madrid, Spain}
   \date{Received 2 January 2019 / Accepted 11 June 2019}

   \abstract
      {Hypervelocity stars move fast enough to leave the gravitational field 
       of their home galaxies and venture into intergalactic space. The most 
       extreme examples known have estimated speeds in excess of 
       1000~km~s$^{-1}$. These can be easily induced at the centres of 
       galaxies via close encounters between binary stars and supermassive 
       black holes; however, a number of other mechanisms operating elsewhere 
       can produce them as well. 
       }
      {Recent studies suggest that hypervelocity stars are ubiquitous in the 
       local Universe. In the Milky Way, the known hypervelocity stars are 
       anisotropically distributed, but it is unclear why. Here, we used 
       {\it{Gaia}} Data Release 2 (DR2) data to perform a systematic 
       exploration aimed at confirming or refuting these findings. 
       }
      {Our basic premise is that the farther the candidate hypervelocity 
       stars are, the more likely they are to be unbound from the Galaxy. We
       used the statistical analysis of both the spatial distribution and 
       kinematics of these objects to achieve our goals. Monte Carlo sampling 
       techniques were applied to deal with large uncertainties. No global 
       parallax zero-point correction was performed.
       }
      {Focussing on nominal Galactocentric distances greater than 30~kpc, 
       which are the most distant candidates, we isolated a sample with 
       speeds in excess of 500~km~s$^{-1}$ that exhibits a certain degree of 
       anisotropy but remains compatible with possible systematic effects. We 
       find that the effect of the Eddington-Trumpler-Weaver bias is 
       important in our case: over 80\% of our sources are probably located 
       further away than implied by their parallaxes; therefore, most of our 
       velocity estimates are lower limits. If this bias is as strong as 
       suggested here, the contamination by disc stars may not significantly 
       affect our overall conclusions.  
       }
      {The subsample with the lowest uncertainties shows stronger, but 
       obviously systematic, anisotropies and includes a number of candidates 
       of possible extragalactic origin and young age with speeds of up to 
       2000~km~s$^{-1}$.
       }

         \keywords{methods: statistical -- celestial mechanics --
                   stars: kinematics and dynamics -- Galaxy: disc --
                   Galaxy: kinematics and dynamics -- Galaxy: structure
                  }

   \maketitle

   \section{Introduction\label{Intro}}
      Hypervelocity celestial objects have characteristic velocities of the order of 1000~km~s$^{-1}$, meaning those travelling 100~kpc in 
      less than 100~Myr, and they may be ubiquitous in the Universe. The existence of hypervelocity stars was first predicted by 
      \citet{1988Natur.331..687H} and later discussed by \citet{2003ApJ...599.1129Y}, but the first bona fide hypervelocity star 
      (HV~1 or SDSS~J090744.99+024506.9 with a heliocentric radial velocity of 853$\pm$12~km~s$^{-1}$) was identified as such by 
      \citet{2005ApJ...622L..33B}. The first hypervelocity star cluster, HVGC-1 or H70848 with an offset from the systemic velocity of 
      the Virgo Cluster $>2300$~km~s$^{-1}$, was found by \citet{2014ApJ...787L..11C}. In addition, the picture emerging from the 
      spectroscopic analysis of large samples of star-forming galaxies in the local Universe hints at a significant number of high-velocity 
      runaway stars and hypervelocity stars that may be following radial trajectories, away from their original hosts (for example, see 
      \citealt{2016A&A...588A..41C}). 

      It is however not easy to separate true hypervelocity stars from the high-velocity tail associated with the stellar halo, both in the 
      Milky Way galaxy \citep{2019MNRAS.485.3514D} and elsewhere. In the following, we will consider that hypervelocity stars are those 
      that are not gravitationally bound to the Milky Way galaxy or to any other galaxy for that matter, though some researchers use the 
      term to refer exclusively to those stars that were ejected after interacting with the supermassive black hole in the Galactic centre 
      (see the review by \citealt{2015ARA&A..53...15B}). An extensive and timely updated source of data on this subject is the Open Fast 
      Stars Catalog\footnote{\url{https://faststars.space/}} \citep{2017ApJ...835...64G,2018MNRAS.479.2789B}. 

      The known hypervelocity stars with a probable origin in the Milky Way galaxy exhibit a statistically significant anisotropic spatial 
      spread \citep{2009ApJ...690L..69B} and it is not well understood why. Five recent studies \citep{2018MNRAS.479.2789B,
      2018ApJ...868...25B,2018ApJ...866..121H,2018A&A...620A..48I,2018MNRAS.tmp.2466M} use {\it Gaia} Data Release 2 (DR2) data 
      \citep{2016A&A...595A...1G,2018A&A...616A...1G} to independently revisit the topic of hypervelocity stars in the Milky Way. Both 
      \citet{2018MNRAS.479.2789B} and \citet{2018A&A...620A..48I} pay particular attention to reanalysing previously known candidates using 
      the new data; meaning that they do not identify new ones in the {\it Gaia} catalogue. 

      \citet{2018MNRAS.479.2789B} conclude that although known early-type hypervelocity stars may have been ejected from the Galactic 
      centre, the vast majority of late-type candidates could be currently bound to the Galaxy and therefore could be part of the 
      high-velocity tail associated with the stellar disc and halo, or perhaps could be debris from disrupted galaxies. 
      \citet{2018ApJ...868...25B} single out a number of promising new candidate hypervelocity stars at distances in the range of 
      10--15~kpc. \citet{2018ApJ...866..121H} focus on identifying nearby, metal-poor, and relatively old extreme velocity candidates. 
      \citet{2018A&A...620A..48I} find that among the known hypervelocity stars, the fraction with an origin in the disc rather than the 
      Galactic centre dominates. \citet{2018MNRAS.tmp.2466M} isolate a sample that includes both candidates of Galactic and extragalactic 
      origin, but they do not find any candidate consistent with an origin in the Galactic centre. However, they find strong evidence for a 
      population of high-velocity stars with a probable origin outside the Milky Way galaxy. This interpretation agrees well with the ideas 
      put forward by \citet{2016A&A...588A..41C} for example. The purportedly fastest star in the {\it Gaia} catalogue, Gaia~DR2 
      5932173855446728064 \citep{2018ApJ...868...25B,2018MNRAS.tmp.2466M}, has been found to be spurious \citep{2019MNRAS.486.2618B}. 

      Most new hypervelocity star candidates identified in {\it Gaia} DR2 are inside or relatively close to the nominal edge of the Milky 
      Way disc, 15~kpc from the Galactic centre. However, stars located well beyond such a distance may have a higher probability of being
      hypervelocity stars and/or having an extragalactic origin. With this working hypothesis in mind, here we used {\it Gaia} DR2 data 
      to perform a systematic exploration of the spatial distribution of distant, nominal Galactocentric distances greater than 30~kpc,
      hypervelocity star candidates. This investigation is aimed at understanding the origin of any putative anisotropic spatial 
      distribution, but also at confirming or refuting the presence of a significant population of high-velocity stars of intergalactic 
      provenance. This paper is organized as follows. Section~\ref{Dat} discusses data selection issues and the overall approach applied in 
      our study. The results of the full sample are presented and discussed in Sect.~\ref{Results}. Section~\ref{BestRes} focusses on the 
      subsample with the lowest uncertainties. In Sect.~\ref{StatSig}, we evaluate the statistical significance of our findings. Results are 
      discussed in Sect.~\ref{Discus} and conclusions are summarized in Sect.~\ref{Concluding}.

   \section{Data selection and processing\label{Dat}}
      The basic premise that guides our exploration using {\it Gaia} DR2 data is that the farther the candidate hypervelocity stars are, the 
      more likely they are to be unbound from the Milky Way galaxy; this is particularly true for candidates probably located well beyond 
      the Galactic centre. This assumption leads us to deal with samples that have been neglected by previously published studies 
      \citep{2018MNRAS.479.2789B,2018ApJ...868...25B,2018ApJ...866..121H,2018A&A...620A..48I,2018MNRAS.tmp.2466M} because the values of 
      their relative parallax errors ($\sigma_{\pi}/\pi$, where $\pi$ is the measured value of the astrometric parallax and $\sigma_{\pi}$, 
      its uncertainty) are inherently large. It is however possible to obtain statistically significant results if the effects of the 
      parallax errors are well characterised and understood when the sample is magnitude-limited (for example, see the discussion in 
      Sect.~3.6 of \citealt{1998gaas.book.....B}). In the following, averages, standard deviations, medians, interquartile ranges (IQRs), 
      and other statistical parameters have been computed in the usual way (for instance, see \citealt{2002nrca.book.....P,
      2012psa..book.....W}). 

      \subsection{The sample and the uncertainties\label{SandU}}
         In order to interpret our results with confidence (by using colour-magnitude diagrams for example), we focussed on those sources 
         with estimated values of the line-of-sight extinction $A_G$ and reddening $E(G_{\rm BP}-G_{\rm RP})$; {\it Gaia} DR2 includes 
         87\,733\,672 such sources,\footnote{\url{https://www.cosmos.esa.int/web/gaia/dr2}} all of them have strictly positive values of the 
         parallax. Out of this sample, 4\,831\,731 sources have positions, parallax, radial velocity, and proper motions. This smaller 
         sample is suitable for an analysis in the Galactocentric frame of reference. 

         We computed Galactocentric positions using the value of the distance between the Sun and the Galactic centre (Sgr A$^{*}$) given by 
         \citet{2019arXiv190405721A}, 8.18~kpc. These positions are in the Galactocentric standard of rest that is a right-handed 
         coordinate system centred at the Galactic centre with positive axes in the directions of the Galactic centre, Galactic rotation, 
         and the North Galactic Pole (NGP) as discussed by \citet{1987AJ.....93..864J} for example. Galactocentric Galactic velocity 
         components were calculated as described by \citet{1987AJ.....93..864J}, considering the values of the Solar motion computed by
         \citet{2010MNRAS.403.1829S} and the value of the in-plane circular motion of the local standard of rest around the Galactic
         centre discussed by \citet{2014ApJ...783..130R}. These values are also used by \citet{2018ApJ...868...25B}. From the 
         smaller sample, our software pipeline produced 15\,681 sources with nominal Galactocentric distances $>30$~kpc and full kinematics 
         with uncertainties; this is our primary full sample.

         As the input data have large uncertainties, we did not use the error expressions presented by \citet{1987AJ.....93..864J} but a 
         Monte Carlo \citep{MU49,2002nrca.book.....P} sampling technique to perform error estimation. The Monte Carlo methodology used here 
         also provides estimates for the most probable values of the various parameters. For each source in our primary sample, we generated 
         10$^{5}$ realizations of its astrometric and photometric parameters to produce lists of ordered data values. Neither mean values 
         nor standard deviations of computed parameters (with the exception of those of input {\it Gaia} DR2 data) are used in this work. As 
         measures of central tendency and dispersion, we used the median (50th percentile) and the 16th and 84th percentiles, respectively. 
         These are the values given in the sections, figures, and tables. The covariance matrix of parallax and proper motions was computed 
         using the mean values, standard deviations, and correlation coefficients provided by {\it Gaia} DR2.

\begin{figure*}
        \centering
         \includegraphics[width=0.49\linewidth]{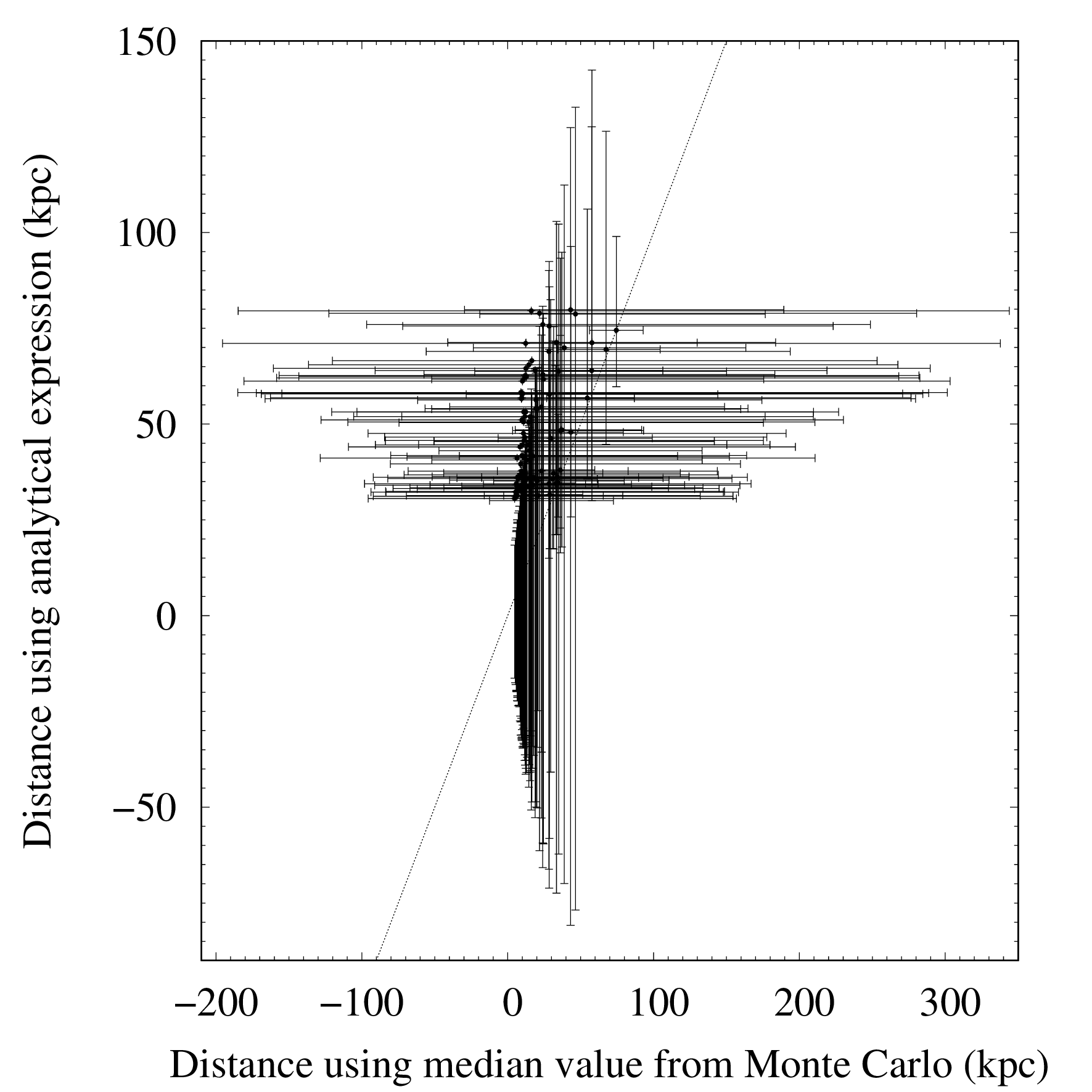}
         \includegraphics[width=0.49\linewidth]{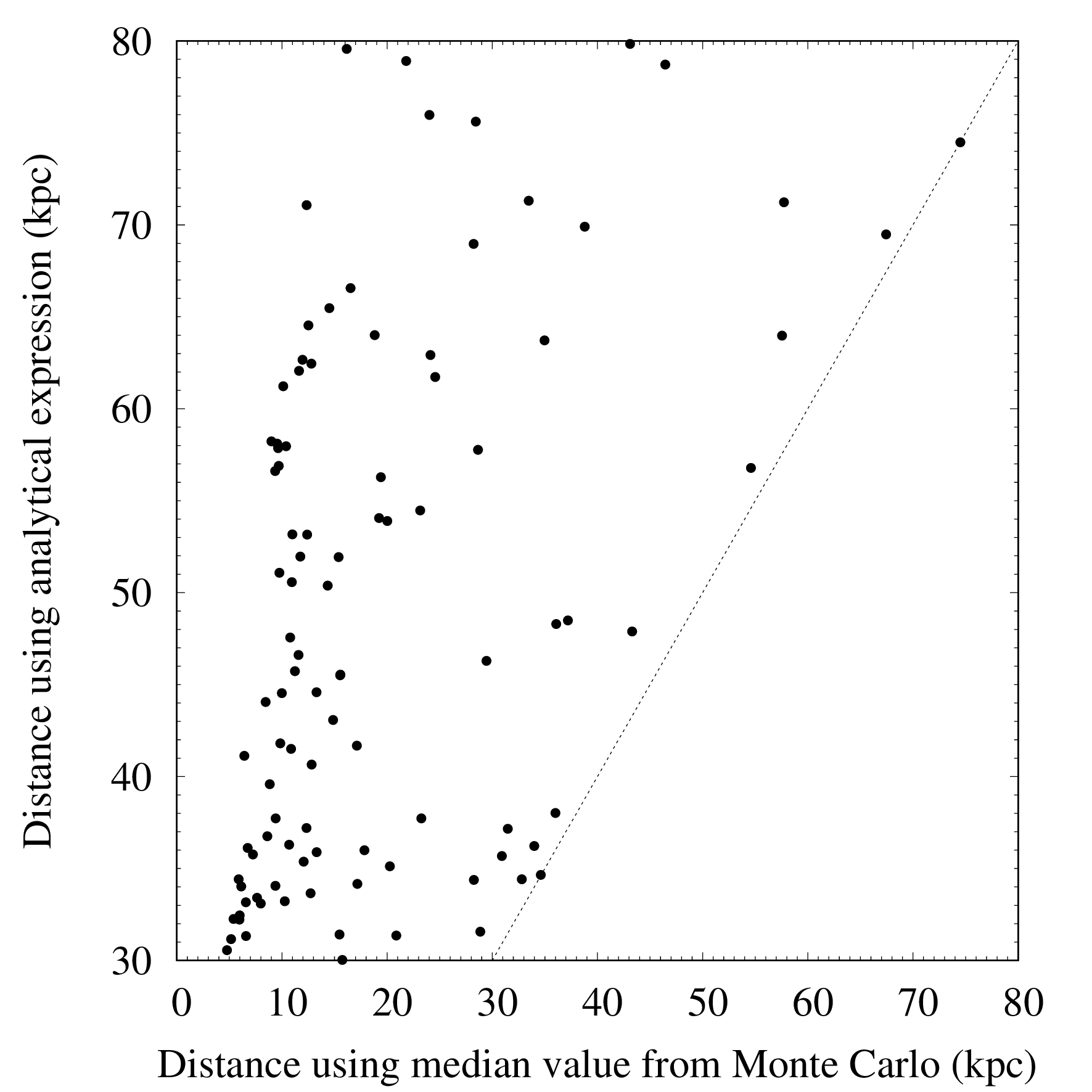}
         \caption{Analytical versus Monte Carlo distances for 100 representative synthetic data with $\sigma_{\pi}/\pi\in[0.2, 4.2]$ and 
                  $\pi\in[0.0125, 0.0333]$~mas (uniformly distributed). These are shown with (left) and without (right) error bars, computed 
                  as described in the text. The discontinuous line shows the diagonal ($d=d_{\rm c}$) for reference.
                 }
         \label{monte}
      \end{figure*}

         For example, if $\vec{r}$ is a vector made of univariate Gaussian random numbers (components $r_{i}$ with $i=1,3$), the required 
         multivariate Gaussian random samples are given by the expressions:
         \begin{equation}
            \begin{aligned}
               \pi_{\rm c}           & = \pi + a_{11}\,r_{1} \\
               \mu_{\alpha\ {\rm c}} & = \mu_{\alpha} + a_{22}\,r_{2} + a_{21}\,r_{1} \\
               \mu_{\delta\ {\rm c}} & = \mu_{\delta} + a_{33}\,r_{3} + a_{32}\,r_{2} + a_{31}\,r_{1} \,,
               \label{cova}
            \end{aligned}
         \end{equation} 
         where $\pi$, $\mu_{\alpha}$, and $\mu_{\delta}$ are the values of absolute stellar parallax and proper motions in right ascension 
         and declination directions provided by {\it Gaia} DR2 and the $a_{ij}$ coefficients are given by:
         \begin{equation}
            \begin{aligned}
               a_{11} & = \sigma_{\pi} \\
               a_{21} & = \rho_{\pi\ \mu_{\alpha}}\,\sigma_{\mu_{\alpha}} \\
               a_{31} & = \rho_{\pi\ \mu_{\delta}}\,\sigma_{\mu_{\delta}} \\
               a_{22} & = \sqrt{\sigma_{\mu_{\alpha}}^{2} - a_{21}^{2}} \\
               a_{32} & = (\rho_{\mu_{\alpha}\ \mu_{\delta}}\,\sigma_{\mu_{\alpha}}\,\sigma_{\mu_{\delta}} - a_{21}\,a_{31}) / a_{22} \\
               a_{33} & = \sqrt{\sigma_{\mu_{\delta}}^{2} - a_{31}^{2} - a_{32}^{2}} \,,
            \end{aligned}
         \end{equation}
         where $\sigma_{\pi}$, $\sigma_{\mu_{\alpha}}$, and $\sigma_{\mu_{\delta}}$ are the standard errors in parallax and proper motions 
         from {\it Gaia} DR2, and $\rho_{\pi\ \mu_{\alpha}}$, $\rho_{\pi\ \mu_{\delta}}$, and $\rho_{\mu_{\alpha}\ \mu_{\delta}}$ their 
         respective correlation coefficients, also from {\it Gaia} DR2. For each $\pi_{\rm c}$, we computed the value of the distance by 
         applying the usual relationship, $d_{\rm c}=1/\pi_{\rm c}$. If the value $\pi_{\rm c}$ drawn from the normal distribution turned 
         out to be negative, it was not discarded. The resulting list of 10$^{5}$ values of $d_{\rm c}$ was ordered, and used to obtain the 
         median and the 16th and 84th percentiles (see Fig.~\ref{monte}). An equivalent approach was used to estimate the relevant values of 
         the percentiles of interest for other parameters such as the Galactocentric distance and Galactic velocity components; the values 
         are quoted in terms of the median of the distribution, with uncertainties derived from the 16th and 84th percentiles. Given the 
         fact that negative values of $\pi_{\rm c}$ were not discarded, the uncertainties were fully characterised because the information 
         on the tail of the probability distribution of the parallax below zero was not lost in the Monte Carlo sampling.

         In order to generate random numbers ($r_{i}$) with a standard Gaussian or normal distribution with mean 0 and standard deviation 1, 
         we applied the Box-Muller method \citep{BM58,2002nrca.book.....P}. The Box-Muller method requires a uniform random variable as 
         seed; when a computer is used to produce a uniform random variable, it will inevitably have some inaccuracies because there is a 
         lower boundary on how close numbers can be to 0. For a 64 bits computer the smallest non-zero number is $2^{-64}$, which means that 
         the Box-Muller method will not produce random variables more than 9.42 standard deviations from the mean (see 
         \citealt{2002nrca.book.....P}).
%
%-------------------------------------------------------------------------------------------------------------------------------------------
%
%
%-------------------------------------------------------------------------------------------------------------------------------------------
%

      \subsection{Parallax zero-point\label{DR2ZP}}
         \citet{2018A&A...616A...2L} use quasars to study the large-scale systematics in the astrometric quality of {\it Gaia} DR2 data, 
         concluding that there is a significantly negative global zero-point, $-0.029$~mas. In other words, the reported {\it Gaia} DR2 
         parallaxes are too small and hence the distances are systematically overestimated if one adopts the usual relationship. In our 
         particular case, this issue might lead to overestimated distances and therefore unrealistically high values of the Galactocentric 
         velocity. A possible solution to this problem could be in using $d_{\rm c}=1/(\pi_{\rm c}+0.029~{\rm mas})$ for the calculations 
         described above. This is however not advisable for a number of reasons.

         On the one hand, the actual value of the global zero-point is controversial. Although \citet{2019ApJ...872...85G} find a best 
         value of the parallax zero-point of $-0.031\pm0.011$~mas, which is fully consistent with the one reported by 
         \citet{2018A&A...616A...2L}, other authors find larger parallax offsets. \citet{2018ApJ...861..126R} give a value of
         $-0.046\pm0.013$~mas, \citet{2019arXiv190208634L} find a value of $-0.052\pm0.002$~mas, \citet{2018arXiv180502650Z} 
         obtain $-0.053\pm0.003$~mas, \citet{2019MNRAS.tmp.1390S} find a value of $-0.054\pm0.006$~mas, \citet{2018A&A...616A..17A} 
         use star clusters to obtain $-0.067\pm0.012$~mas (see their Fig.~16), \citet{2019ApJ...875..114X} compute a parallax offset of 
         $-0.075\pm0.029$~mas, and \citet{2018ApJ...862...61S} find a value of $-0.082\pm0.033$~mas. The wide range in the values 
         proposed for this offset mainly arises from the different types of objects (and thus different colour and magnitude intervals) 
         used to compute the various estimates. Although all the proposed values of the global zero-point are below 0.100~mas in absolute 
         terms (which is the pre-launch, expected global systematics), it must be emphasized that they are consistently negative and below 
         the one determined by \citet{2018A&A...616A...2L}, $-0.029$~mas. Our sample has nominal parallaxes $<0.033$~mas; if we perform 
         this correction, even if we use the most optimistic value, $-0.029$~mas, the size of the resulting sample would become virtually 
         zero (see Appendix \ref{ZPoffset}). 

         On the other hand, \citet{2018A&A...616A..17A} show that the parallax offset is partly dependent on the scanning pattern 
         of the {\it Gaia} mission, which makes it a function of the coordinates (see their Fig.~15). Both \citet{2018A&A...616A...2L} 
         and \citet{2018A&A...616A..17A} explicitly discourage a global zero-point correction, particularly in cases (like ours) where 
         the sample is not well distributed over the entire sky (see below). For these reasons, we do not correct the parallaxes for the 
         global zero-point.

      \subsection{Selection criteria\label{SeCr}}
         Our identification of distant hypervelocity star candidates was based on the values of Galactocentric distance and velocity. At
         Galactocentric distances $>30$~kpc, a Galactocentric velocity $>500$~km~s$^{-1}$ is probably signalling the presence of robust
         hypervelocity star candidates or at least members of the high-velocity tail associated with the stellar halo 
         ---\citet{2019MNRAS.485.3514D} find a value for the escape speed in the neighbourhood of the Sun of 
         $528^{+24}_{-25}$~km~s$^{-1}$, therefore the value at distances greater than 30~kpc from the Galactic centre is expected to be 
         lower. These criteria are more conservative than the ones used recently to discuss two new hypervelocity stars from the LAMOST 
         spectroscopic surveys \citep{2017ApJ...847L...9H}. 

         As for the origin of the candidates, we used the angle between the position and velocity vectors as a flag (this technique is 
         discussed by \citealt{2018ApJ...868...25B}). A hypervelocity star coming from the region of the Galactic centre must have a value 
         of the angle close to 0{\degr} as both vectors are nearly parallel, bound stars or those moving tangentially are expected to have 
         values of the angle in the neighbourhood of 90{\degr} as both vectors are nearly perpendicular, and those stars coming straight 
         from intergalactic space and headed for the region of the Galactic centre should have a value of the angle close to 180{\degr} when 
         both vectors are almost antiparallel, but in general $>90$\degr. However, the angle criterion has some weaknesses. A fast-moving 
         star coming from outside the Milky Way galaxy may have experienced a hyperbolic encounter with the Galactic centre and the angle 
         between the position and velocity vectors could be close to 0{\degr} when moving away from the Galactic centre and back into 
         intergalactic space. On the other hand, any intergalactic star traversing the Galaxy without experiencing significant deflection 
         will mimic an origin within the disc if located well outside the Galactic bulge. However, even in such cases, the estimated age of 
         the candidate may help in discarding (or not) a possible intergalactic provenance.

\begin{figure*}
        \centering
         \includegraphics[width=0.49\linewidth]{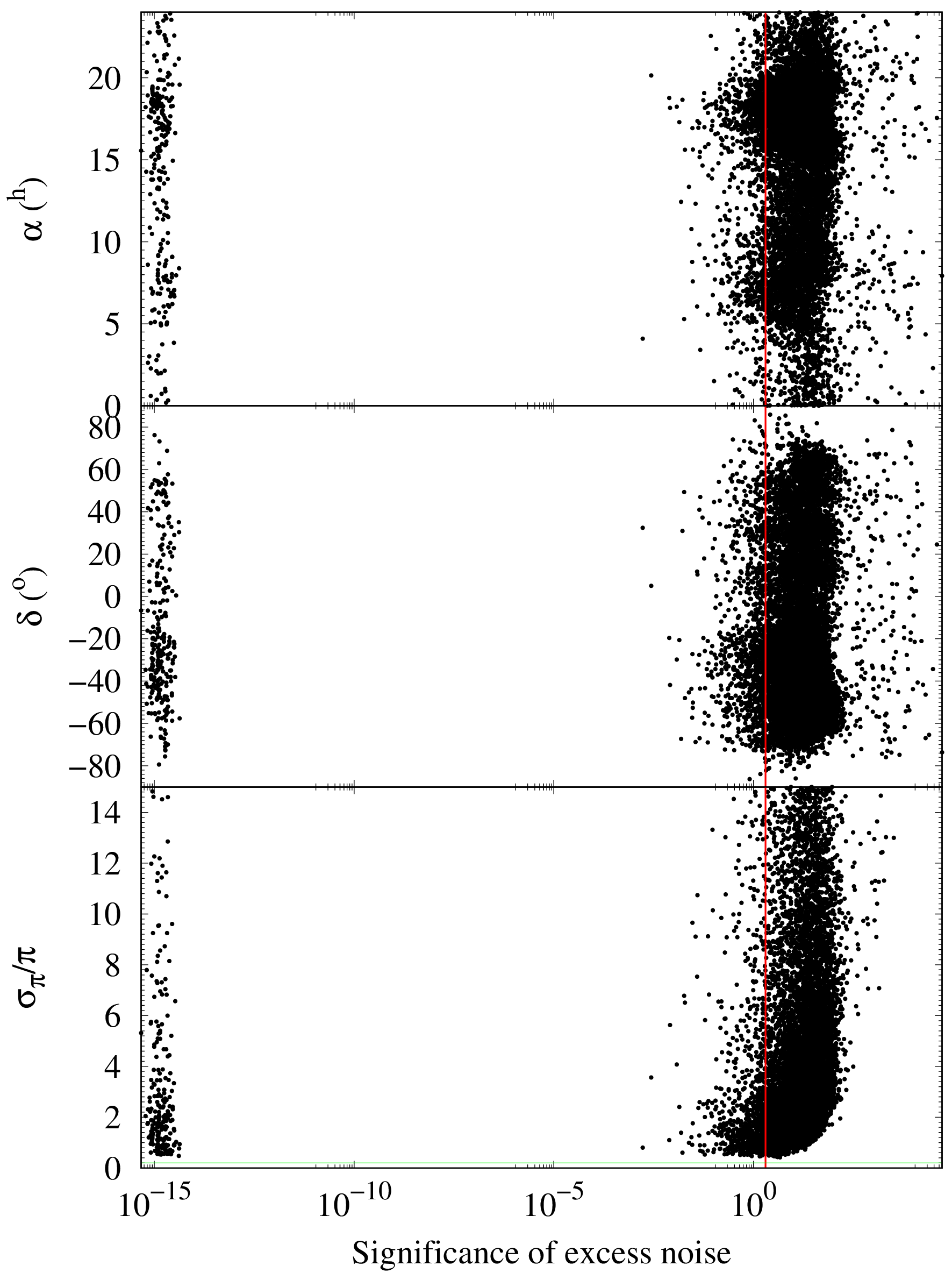}
         \includegraphics[width=0.49\linewidth]{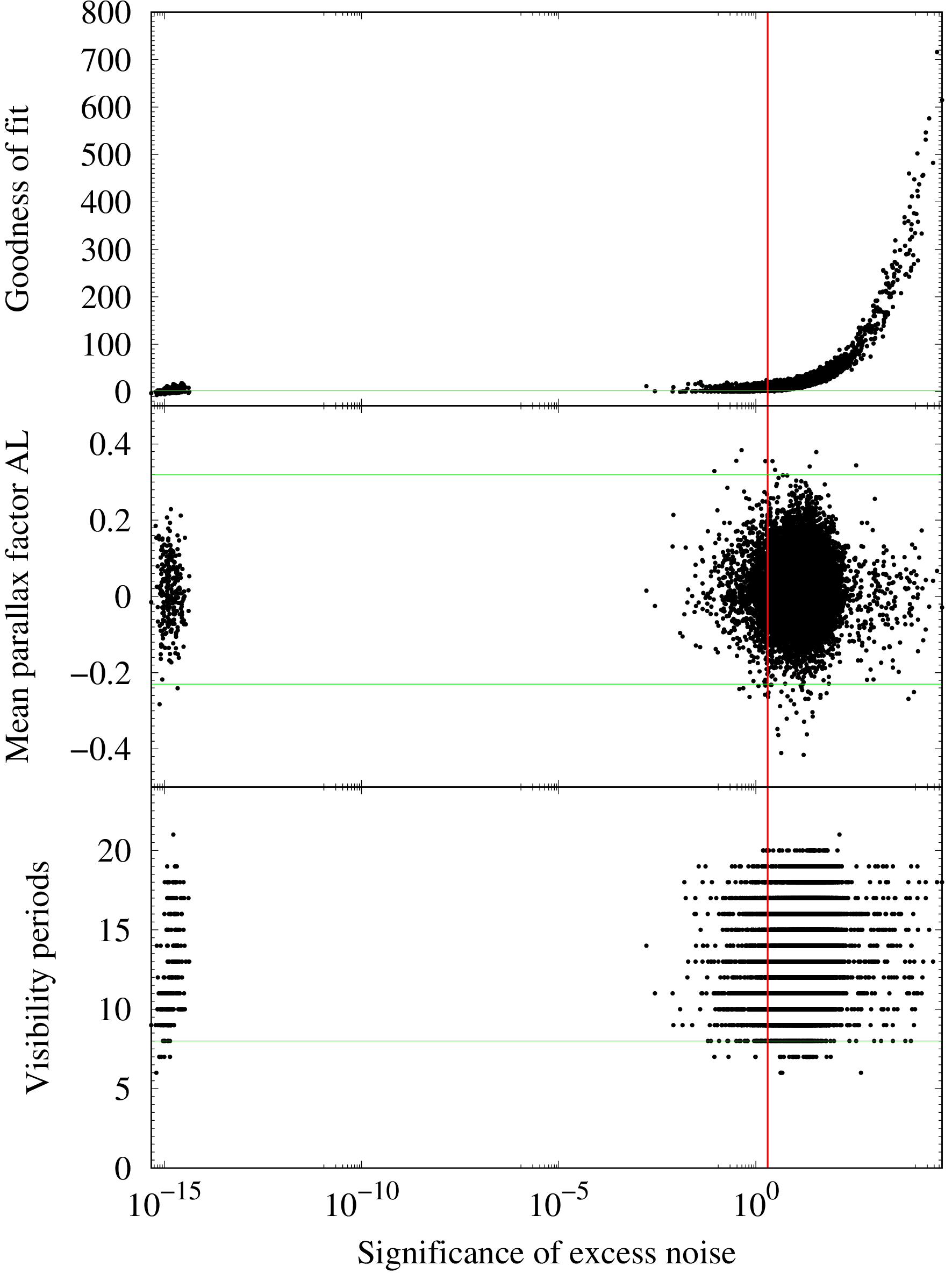}
         \caption{Quality summary of the primary sample (15\,681 sources). Position and relative parallax error as a function of the 
                  parameter {\tt astrometric\_excess\_noise\_sig} or significance of excess noise (left-hand side panels). 
                  \citet{2018A&A...616A...2L} indicate that astrometrically well-behaved sources have a significance of excess noise 
                  $\leq2$ (region on the left of the red line). The green line in the left-hand, bottom panel signals 
                  $\sigma_{\pi}/\pi=0.2$. Dependency between quality flags (right-hand side panels). 
                  \citet{2018A&A...616A...2L} state that sources with {\tt astrometric\_gof\_al} or goodness of fit $<3$ (below the 
                  green line of the top panel), {\tt mean\_varpi\_factor\_al} or mean parallax factor AL in the interval $[-0.23, 0.32]$ 
                  (between the green lines of the middle panel), and {\tt visibility\_periods\_used} or visibility periods $>8$ (above the 
                  green line of the bottom panel) should be used. Not shown here, but also applied, is {\tt rv\_nb\_transits} or number of
                  observations used to compute the value of the radial velocity $>5$.
                 }
         \label{quality}
      \end{figure*}

         The sample of high-velocity stars may be contaminated by stars with poor astrometric solutions. In order to single out those 
         sources with reliable astrometry, we applied the criteria discussed by \citet{2018A&A...616A...2L} and 
         \citet{2018MNRAS.tmp.2466M} (regarding the parameters {\tt astrometric\_gof\_al}, {\tt astrometric\_excess\_noise\_sig},
         {\tt mean\_varpi\_factor\_al}, {\tt visibility\_periods\_used}, and {\tt rv\_nb\_transits}). The data and the criteria are 
         presented in Fig.~\ref{quality}. After the application of these criteria, the primary sample of 15\,681 was reduced to 393 stars, 
         174 of them have median Galactocentric distance greater than 30~kpc and median Galactocentric Galactic velocity greater than 
         500~km~s$^{-1}$ (computed using the Monte Carlo methodology described above).
%
%-------------------------------------------------------------------------------------------------------------------------------------------
%
%
%-------------------------------------------------------------------------------------------------------------------------------------------
%

      \subsection{Extinction and reddening\label{ExandRed}}
         Our {\it Gaia} DR2 sources have estimated values of the line-of-sight extinction $A_G$ and reddening $E(G_{\rm BP}-G_{\rm RP})$.
         These values are highly biased due to the model grids that were used to train the machine learning algorithms employed and the
         extinction was computed considering the usual relationship between parallax and distance, without taking into account the parallax 
         error or parallax zero-point offset \citep{2018A&A...616A...1G}. \citet{2018A&A...616A...8A} indicate that the typical 
         accuracy in $A_G$ is of the order of 0.46~mag and that of $E(G_{\rm BP}-G_{\rm RP})$ could be 0.23~mag. Classical sources to obtain
         estimates of Galactic dust reddening and extinction are \citet{1998ApJ...500..525S} and \citet{2011ApJ...737..103S}. Their
         datasets are available from the NASA/IPAC InfraRed Science Archive\footnote{\url{https://irsa.ipac.caltech.edu/applications/DUST/}}
         and we retrieved the relevant values to compare with those given by {\it Gaia} DR2 (although the wavelengths are different).

\begin{figure*}
        \centering
         \includegraphics[width=0.49\linewidth]{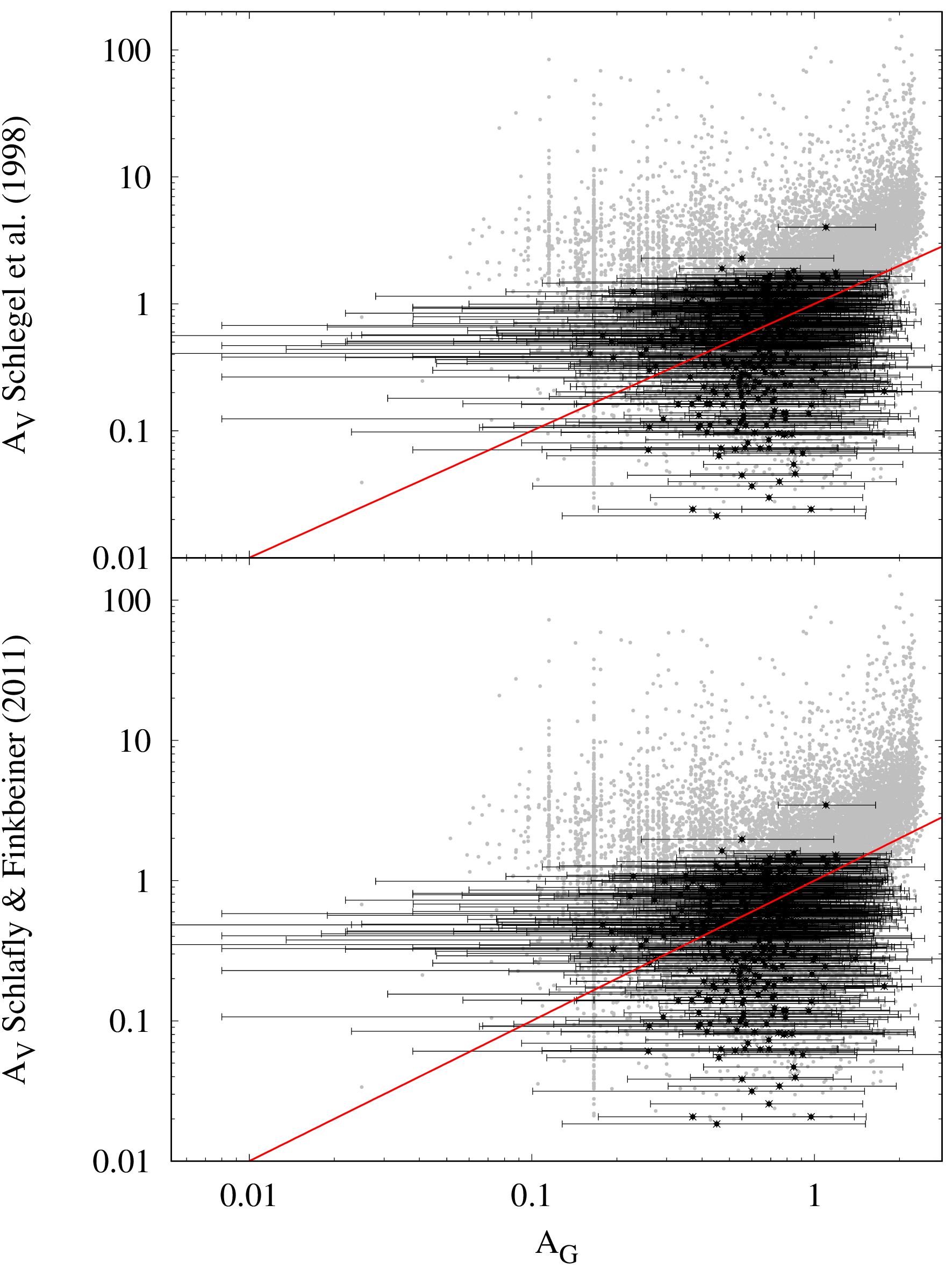}
         \includegraphics[width=0.49\linewidth]{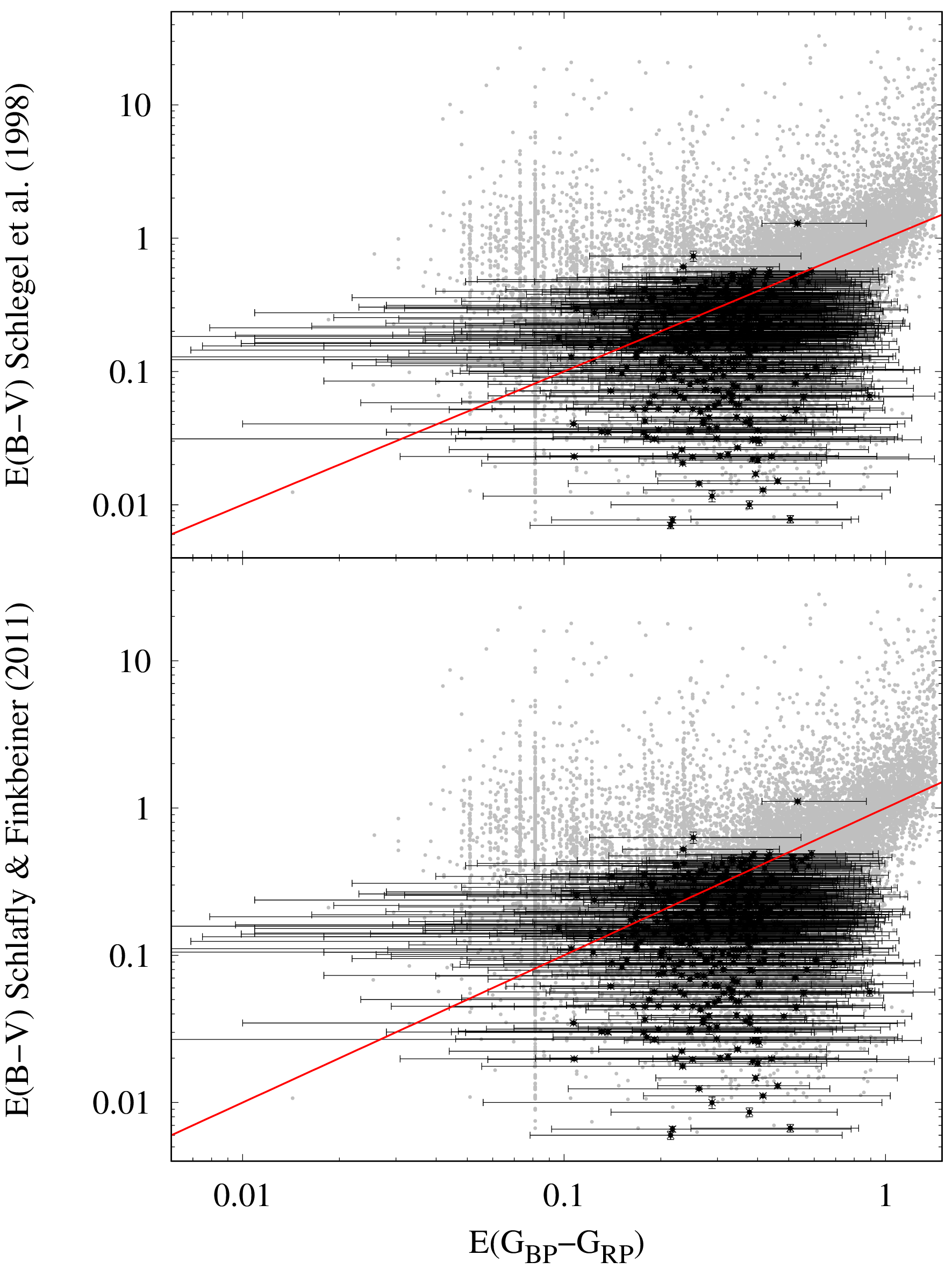}
         \caption{Extinction and reddening, a comparison. The left-hand side panels show extinction values $A_V$ from 
                  \citet{1998ApJ...500..525S}, top panels, and \citet{2011ApJ...737..103S}, bottom panels, versus the line-of-sight 
                  extinction $A_G$ from {\it Gaia} DR2 for the 15\,681 sources in our primary sample (in grey) and the 393 stars with the 
                  best astrometric solutions (in black). The right-hand side panels show colour excesses $E({\rm B}-{\rm V})$ from 
                  \citet{1998ApJ...500..525S} and \citet{2011ApJ...737..103S} versus $E(G_{\rm BP}-G_{\rm RP})$ from {\it Gaia} DR2 
                  for the same sources. The reference line in red is the diagonal. See the text for additional details.
                 }
         \label{extinction}
      \end{figure*}

         Figure~\ref{extinction} is a comparison between the extinction (left-hand side panels) and reddening (right-hand side panels) 
         values from {\it Gaia} DR2 and those derived by \citet{1998ApJ...500..525S}, top panels, and \citet{2011ApJ...737..103S},
         bottom panels, for the 15\,681 sources in our primary sample (in grey) and the 393 stars with the best astrometric solutions (in 
         black). The values derived by \citet{1998ApJ...500..525S} and \citet{2011ApJ...737..103S} are linearly correlated, with 
         those from \citet{1998ApJ...500..525S} being somewhat higher as seen from the reference red diagonal line, meaning 
         $E(G_{\rm BP}-G_{\rm RP})=E({\rm B}-{\rm V})$. The error bars (only for 393 sources) of {\it Gaia} DR2 extinction and reddening 
         show the 16th and 84th percentiles, respectively, those of \citet{1998ApJ...500..525S} and \citet{2011ApJ...737..103S} 
         display the standard deviations, which are very small for reddening and zero for extinction. 
%
%-------------------------------------------------------------------------------------------------------------------------------------------
%
%
%-------------------------------------------------------------------------------------------------------------------------------------------
%

         Although the wavelengths are different and the uncertainties in the {\it Gaia} DR2 values are large, the agreement is fairly good
         for the bulk of sources with the best astrometric solutions although {\it Gaia} DR2 extinction and reddening values tend to be 
         larger than those from \citet{1998ApJ...500..525S} and \citet{2011ApJ...737..103S}. The two-dimensional maps of 
         \citet{1998ApJ...500..525S} and \citet{2011ApJ...737..103S} were obtained without considering any distance information, while
         the {\it Gaia} DR2 extinction values do take into account the distance derived from the usual relationship between parallax and 
         distance, and neglecting its uncertainty. We did not observe, among the 393 stars with the best astrometric solutions, significant 
         numbers of outliers with abnormally low or high values of extinction and reddening; therefore, it seems unlikely that our 
         conclusions could be affected by erroneous extinction and reddening values from {\it Gaia} DR2.  

      \subsection{Are they really that far away?\label{FarAway}}
         In principle, the central problem with our analysis is in the use of sources with large relative parallax errors, larger than those
         discussed in the five recent studies that use {\it Gaia} DR2 data \citep{2018MNRAS.479.2789B,2018ApJ...868...25B,
         2018ApJ...866..121H,2018A&A...620A..48I,2018MNRAS.tmp.2466M}. This choice, which is the result of the basic premise pointed out 
         above, has two main side effects: distances could be both incorrect and underestimated, and their uncertainties cannot be properly 
         computed. 

         On the one hand, it is a well-known fact that when $\sigma_{\pi}/\pi\geq0.2$ it becomes very difficult to estimate the true value 
         of the astrometric parallax (see \citealt{1953stas.book.....T,2015PASP..127..994B}); in other words, the value of the 
         distance obtained by applying the usual relationship, $d=1/\pi$, as well as its associated error, $\Delta{d}=\sigma_{\pi}/\pi^2$, 
         become very unreliable. Within the context of {\it Gaia} DR2 data, our self-imposed lower limit for the nominal Galactocentric 
         distance automatically implies that our primary full sample has $\sigma_{\pi}/\pi>0.2$. On the other hand, the 
         Eddington-Trumpler-Weaver bias \citep{1913MNRAS..73..359E,1940MNRAS.100..354E,1953stas.book.....T} affects parallax-limited samples 
         such as ours. The circumstances that lead to the Eddington-Trumpler-Weaver bias have been revisited multiple times (see 
         \citealt{1973PASP...85..573L,1998MNRAS.294L..41O,2003MNRAS.338..891S,2014MNRAS.444L...6F}) and it is more commonly discussed under 
         the term Lutz-Kelker bias; it is still unclear whether or not it can be corrected (and if it can, how, see 
         \citealt{1973PASP...85..573L,1992MNRAS.256...65K,2014MNRAS.444L...6F}). It is however widely accepted that it causes measured 
         parallaxes to be too large, leading to underestimated distances, when the uncertainty in the measured value becomes a significant 
         fraction of the measurement itself. In our case, this implies that the actual distance to the source is very probably larger than 
         the one computed using the conventional expression, $d=1/\pi$.

         The analysis presented by \citet{1953stas.book.....T} suggested that any values of the distance estimated using $d=1/\pi$ for a
         magnitude-limited sample with $\sigma_{\pi}/\pi>0.2$ could be systematically underestimated, but it did not provide a procedure to 
         obtain the true values. At this point, one may argue that the correct approach to estimate the true distances should involve the 
         application of Bayesian inference, adopting a prior distribution for distances in our Galaxy (see \citealt{2015PASP..127..994B,
         2016ApJ...832..137A,2018AJ....156...58B,2018A&A...616A...9L}). This is particularly important for the stars in our primary full 
         sample, since they are thought to be distant ($>30$~kpc), faint stars with large uncertainties on parallax. 
         \citet{2018AJ....156...58B} compute the distances to 1.33 billion stars in {\it Gaia} DR2 using Bayesian inference and 
         all the sources in our primary full sample have estimated distances in this catalogue. However, the fraction of distant stars in 
         \citet{2018AJ....156...58B} is very negligible, with 31\,298 sources with Heliocentric distance $>15$~kpc, 49 with distance 
         $>25$~kpc, and 4 with distance $>30$~kpc; in other words, most sources with values of the parallax under 0.033~mas in {\it Gaia} 
         DR2 have Bayesian distance estimates in \citet{2018AJ....156...58B} well below 15~kpc. The distance estimates in 
         \citet{2018AJ....156...58B} are biased to lower values if the parallax data are poor because they adopt a small scale length for 
         their prior. This fact interpreted within the context of the Eddington-Trumpler-Weaver bias discussed above led us to enquire how 
         reliable were the estimates provided by \citet{2018AJ....156...58B} and (which is far more important) how strong was the effect 
         of the Eddington-Trumpler-Weaver bias on our sample.   

         Eclipsing binaries are routinely used to find the distances to galaxies both in the Local Group and beyond as they provide an 
         independent and precise alternative to astrometric parallaxes (for example, see \citealt{1991MNRAS.250..119B}). In order to measure 
         the strength of the Eddington-Trumpler-Weaver or Lutz-Kelker bias on our sample, we retrieved results from recent eclipsing 
         binary surveys in the Magellanic Clouds. Using eclipsing binaries, \citet{2013Natur.495...76P} found a value for the 
         distance to the Large Magellanic Cloud (LMC) of 49.97$\pm$0.19 (statistical) $\pm$1.11 (systematic)~kpc (or a parallax of 
         0.0208$\pm$0.0005~mas); also observing eclipsing binaries, \citet{2005MNRAS.357..304H} found a distance of 60.6$\pm$1.0 
         (statistical) $\pm$2.8 (systematic)~kpc (or a parallax of 0.0165$\pm$0.0010~mas) to the Small Magellanic Cloud (SMC). This distance 
         interval covered reasonably well the range of interest for our primary full sample, which is 30--60~kpc for candidates with 
         relative parallax error $<1$ (see Sect.~\ref{BestRes}). These values (of $d$ and its associated $\pi$) can be used as the true ones 
         for all practical purposes as their uncertainties are far lower than those from astrometric parallaxes. In addition, 
         \citet{2014MNRAS.443..432M} provide an extensive catalogue of 1768 eclipsing binaries in the LMC and the one in 
         \citet{2013AcA....63..323P} include 6138 eclipsing binaries in the SMC. The vast majority of these stars have counterparts in 
         {\it Gaia} DR2 and, therefore, measured parallaxes. These values can be used to evaluate both the reliability of the relevant 
         results in \citet{2018AJ....156...58B} within the context of our study and the effect of the Eddington-Trumpler-Weaver bias on 
         our sample. 

\begin{figure*}
        \centering
         \includegraphics[width=0.49\linewidth]{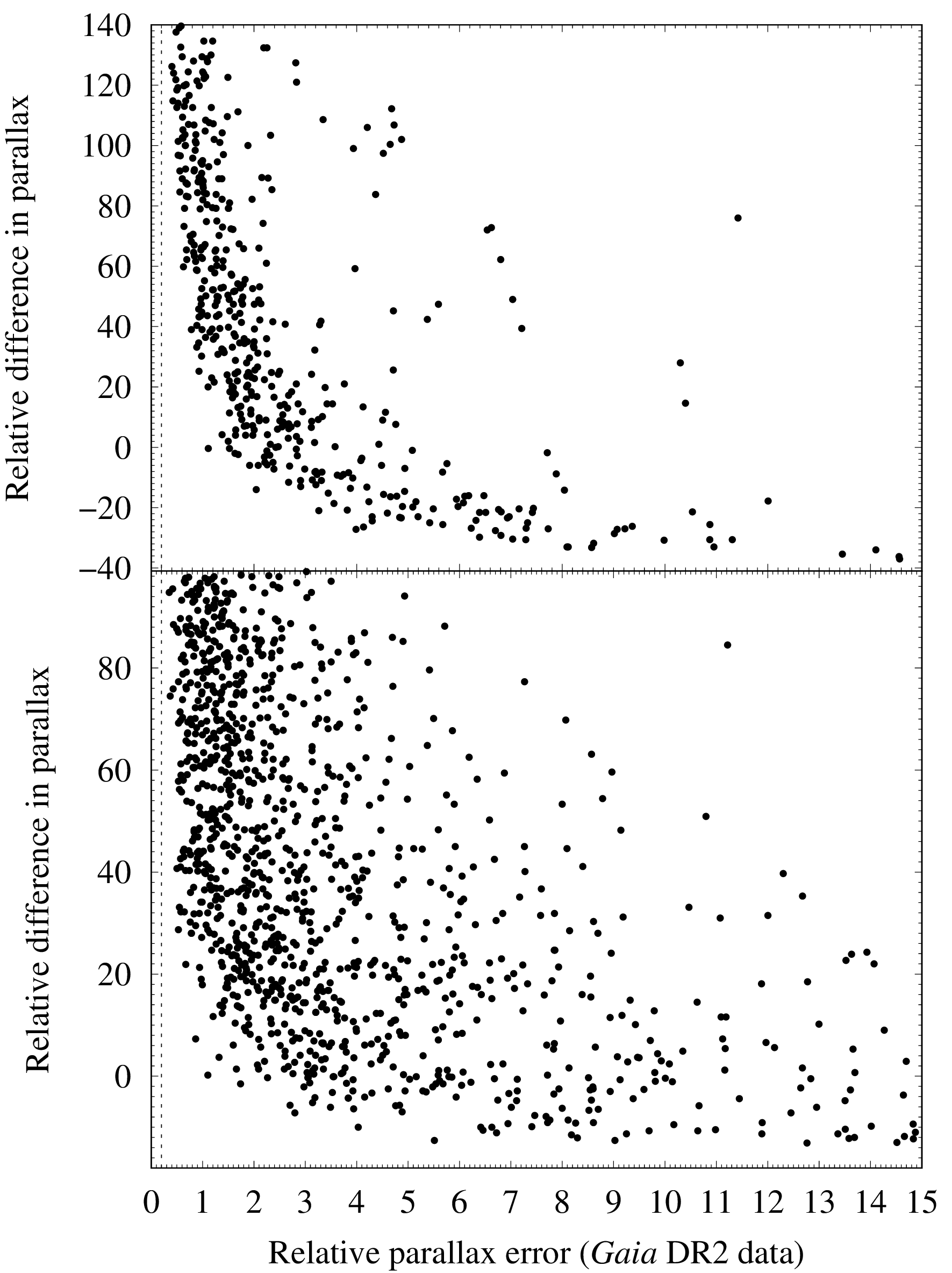}
         \includegraphics[width=0.49\linewidth]{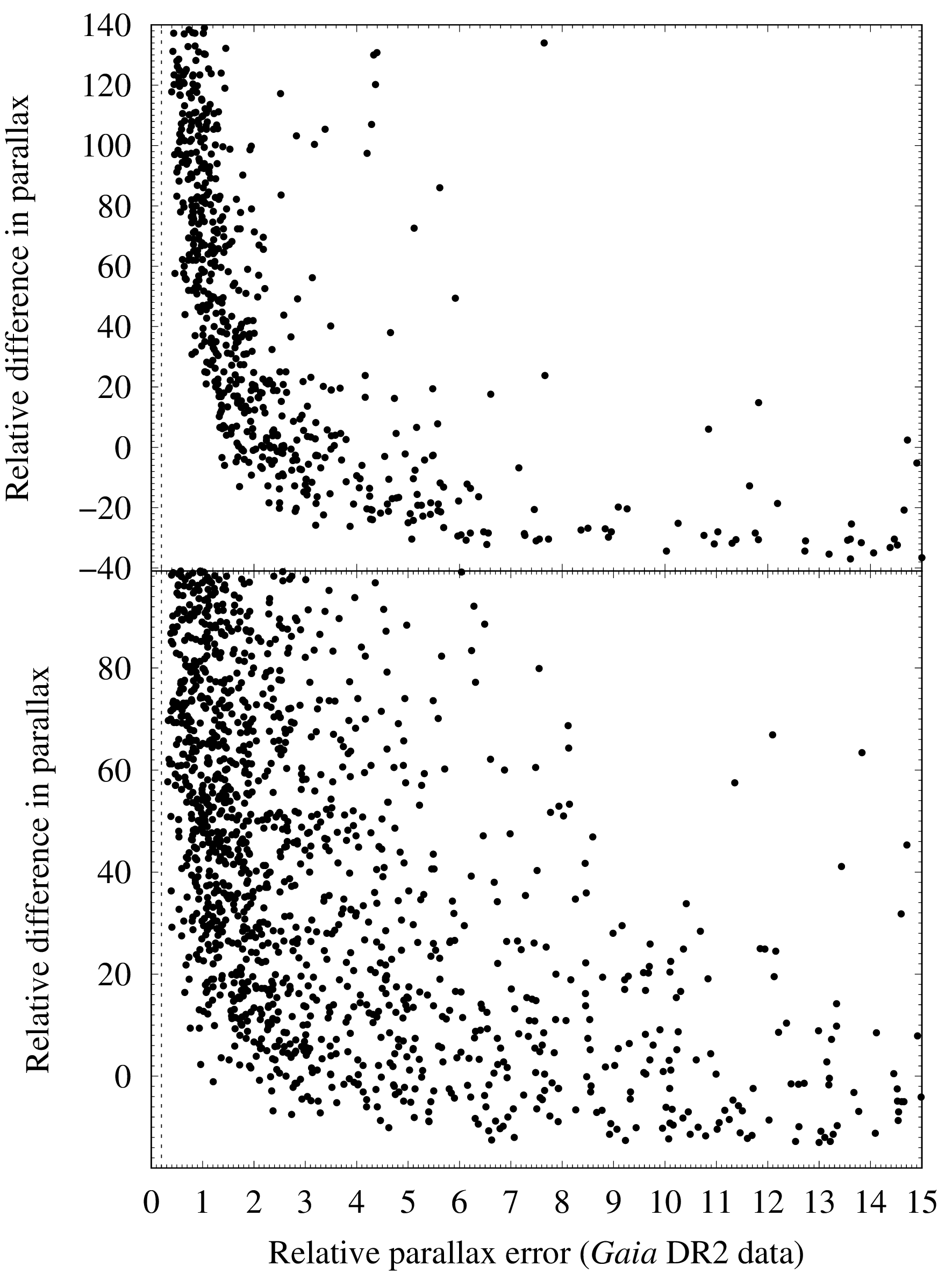}
         \caption{Eddington-Trumpler-Weaver bias. Evaluation of this bias for data from \citet{2014MNRAS.443..432M}, top panel, and 
                  \citet{2013AcA....63..323P}, bottom panel. The $y$-axis shows the difference between the value of the parallax from
                  {\it Gaia} DR2 and the one derived from the assumed distance ---0.0208$\pm$0.0005~mas for the LMC (top panel, 
                  \citealt{2013Natur.495...76P}) and 0.0165$\pm$0.0010~mas for the SMC (bottom panel, \citealt{2005MNRAS.357..304H})--- 
                  divided by the quoted uncertainty (0.0005~mas, top panel, and 0.0010~mas, bottom panel). The $x$-axis shows the 
                  corresponding value of the relative error in parallax from {\it Gaia} DR2. Only sources with strictly positive values of
                  the parallax are plotted; a sample of the full catalogues is shown as the full ranges in $x$ and $y$ are wider. The 
                  condition $\sigma_{\pi}/\pi=0.2$ is plotted as a discontinuous line. The left-hand side panels show results without 
                  considering any global zero-point correction; the right-hand side panels include the correction $(\pi+0.029~{\rm mas})$
                  discussed in the text. 
                 }
         \label{ETWbias}
      \end{figure*}

         Figure~\ref{ETWbias} shows the relative differences in the values of the parallax as a function of their relative errors for 
         sources in \citet{2014MNRAS.443..432M} (top panel, LMC) and \citet{2013AcA....63..323P} (bottom panel, SMC) with strictly 
         positive values of the parallax in {\it Gaia} DR2. The search for matching sources was carried out using the tools provided by 
         VizieR \citep{2000A&AS..143...23O} with a radius of 0\farcs8. The $x$-axis shows the value of $\sigma_{\pi}/\pi$ from {\it Gaia} 
         DR2, while the $y$-axis shows the difference between measured and true parallax divided by the uncertainty in the true parallax 
         (meaning that for an eclipsing binary in the LMC, top panel in Fig.~\ref{ETWbias}, 0.0208~mas was subtracted from the value of 
         the {\it Gaia} DR2 parallax and divided by 0.0005~mas). Positive $y$-values correspond to sources with measured parallax value 
         larger than the true one (or empirical distance lower than the real one). For sources in the LMC with strictly positive values of 
         the parallax in {\it Gaia} DR2, the probability of measuring a value of the parallax greater than the true value is 0.83 (in other 
         words, the probability of having a false negative if the null hypothesis is that the source is at or beyond the LMC), for the SMC 
         we found an equivalent probability of 0.94. Within a frequentist framework and making a simple extrapolation, our interpretation of 
         these results is that the Eddington-Trumpler-Weaver bias is present in our sample, it is very important, and over 80\% of our 
         sources may have significantly underestimated values of their distances when the usual expression, $d=1/\pi$, is applied to convert 
         {\it Gaia} DR2 parallaxes into distances. This bias may be even more important for sources located farther away (compare top and 
         bottom panels in Fig.~\ref{ETWbias}). 
%
%-------------------------------------------------------------------------------------------------------------------------------------------
%
%
%-------------------------------------------------------------------------------------------------------------------------------------------
%

         On the other hand, we repeated the analysis for several other dwarf galaxies (closer than the LMC) and obtained consistent 
         results (virtually 100\% false negatives, albeit using samples of eclipsing binaries made of just a handful of sources). If the 
         global zero-point correction discussed above ---$\pi+0.029~{\rm mas}$ instead of $\pi$, meaning that some sources originally with 
         $\pi<0$ got positive values--- is used, the effect of the Eddington-Trumpler-Weaver bias increases only slightly (see 
         Fig.~\ref{ETWbias}, right-hand side panels). The probability of having a false negative if the null hypothesis is that the source 
         is at or beyond the LMC was 0.827 without the correction and it becomes 0.839 with the correction; for the SMC the corresponding 
         probability goes from 0.939 to 0.943. The papers describing the samples of eclipsing binaries used in our analysis state that the 
         fraction of foreground sources, meaning false positives, in their catalogues is negligible; therefore, the values of the 
         probability can be considered as sufficiently reliable. In this analysis we applied a global zero-point correction to a 
         relatively small region of the sky, although both \citet{2018A&A...616A...2L} ---see their Fig.~13--- and 
         \citet{2018A&A...616A..17A} explicitly discourage this practice. 

\begin{figure}
        \centering
         \includegraphics[width=\linewidth]{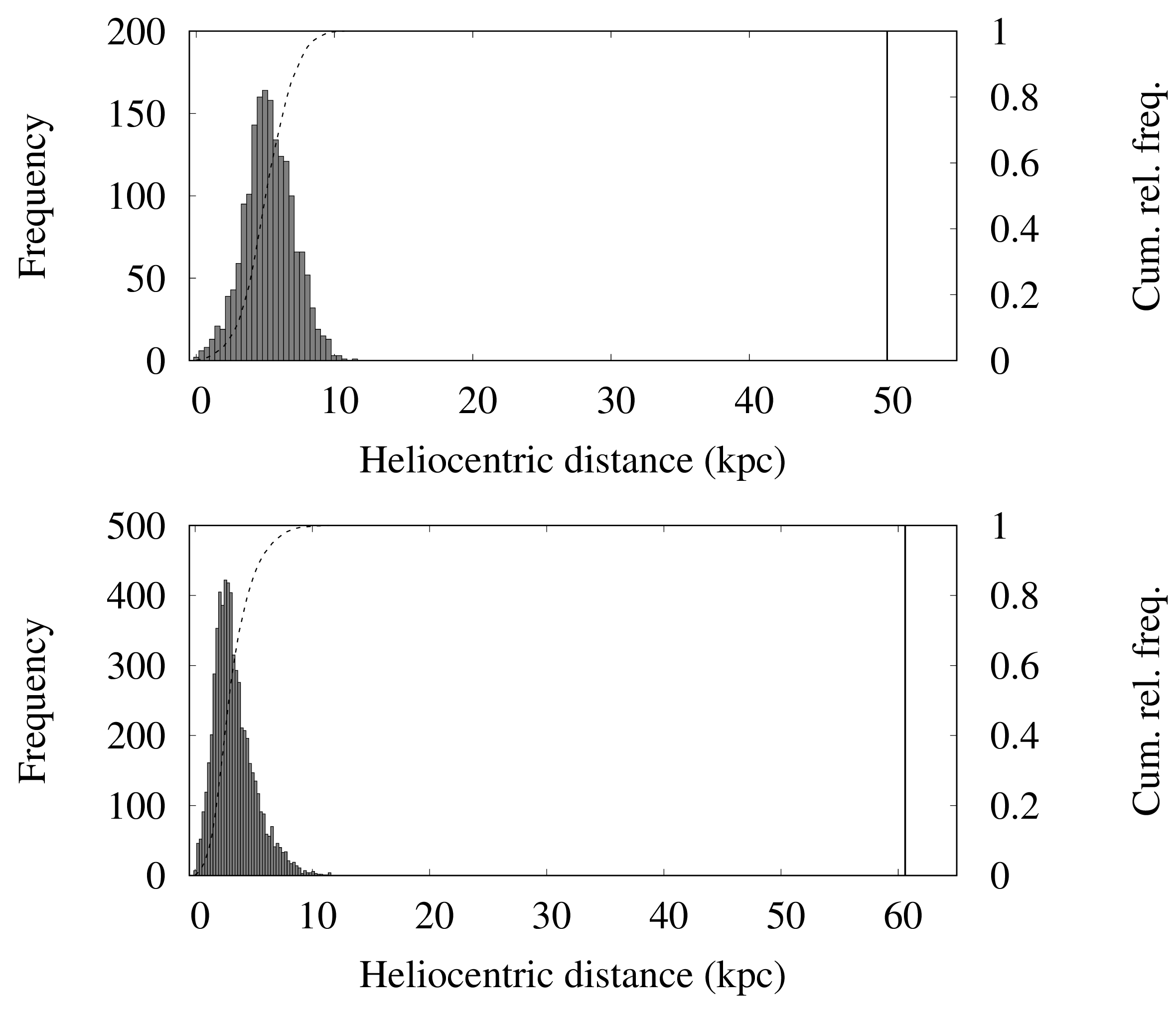}
         \caption{Distribution of Heliocentric distances. Results from \citet{2018AJ....156...58B} for sources in 
                  \citet{2014MNRAS.443..432M}, top panel, and \citet{2013AcA....63..323P}, bottom panel; the respective bin widths of 
                  the histograms are 0.38~kpc and 0.23~kpc (see the text for details). The vertical black bars signal the true values from 
                  \citet{2013Natur.495...76P}, top panel, and \citet{2005MNRAS.357..304H}, bottom panel. The cumulative relative 
                  frequency is plotted as a discontinuous curve.
                 }
         \label{bayes}
      \end{figure}

         In order to evaluate the reliability of the relevant results in \citet{2018AJ....156...58B}, we matched sources as pointed
         out above and constructed histograms with statistically meaningful bin sizes; the bin width was computed using the 
         Freedman-Diaconis rule \citep{FD81}: $2\ {\rm IQR}\ n^{-1/3}$, where $n$ is the number of sources. Figure~\ref{bayes} shows the
         distributions of Heliocentric distances from \citet{2018AJ....156...58B} and the true distances (as vertical black bars). These 
         results indicate that the values computed by \citet{2018AJ....156...58B} cannot be used in our case; in general, distance values
         in \citet{2018AJ....156...58B} are useful in modelling the Galactic global structure by using many stars, but they are not well 
         suited for analyzing individual halo stars when the parallax error is large.
%
%-------------------------------------------------------------------------------------------------------------------------------------------
%
%
%-------------------------------------------------------------------------------------------------------------------------------------------
%

         In summary, we find strong statistical evidence for an affirmative answer to the question posed in the title of this section, at 
         least for most of the sources in the sample selected in Sect.~\ref{SeCr}. This is particularly true for the 174 stars with median 
         Galactocentric distance greater than 30~kpc, median Galactocentric Galactic velocity greater than 500~km~s$^{-1}$, and well-behaved 
         astrometric solutions.

      \subsection{Data processing pipeline validation\label{Pipeline}}
         Our selection criteria are quite different from those used by \citet{2018ApJ...868...25B}, \citet{2018ApJ...866..121H}, or 
         \citet{2018MNRAS.tmp.2466M} and we may have very few sources, if any, in common with them. Nonetheless and in order to assess 
         the quality of our data processing techniques, we computed both Galactocentric distance and velocity for two stars discussed 
         by \citet{2018MNRAS.tmp.2466M}, {\it Gaia} DR2 4326973843264734208 and 4395399303719163904 (they are part of our 4\,831\,731 
         sources sample), in their Table~2 (see Table~\ref{valid}); our results for these stars are shown in Table~\ref{valid} together with
         the results for {\it Gaia} DR2 4326973843264734208 obtained by \citet{2018ApJ...868...25B} and they are statistically consistent 
         with theirs. We also found that the angle between the position and velocity vectors of {\it Gaia} DR2 4326973843264734208 is 
         91\degr, which suggests an extragalactic origin if unbound as position and velocity directions are nearly perpendicular; this is 
         also the interpretation favoured by \citet{2018MNRAS.tmp.2466M}. As for {\it Gaia} DR2 4395399303719163904, we found an angle of 
         52\degr, which is compatible with an origin within the bulk of the Milky Way, but not at the centre, also matching the conclusion in 
         \citet{2018MNRAS.tmp.2466M}. Two other quality-control stars are shown in Table~\ref{valid}. Therefore, our data processing 
         techniques are reliable enough to reproduce results previously obtained by other authors. In any case, the $\sigma_{\pi}/\pi$ 
         values of {\it Gaia} DR2 1383279090527227264, 4326973843264734208, 4395399303719163904, and 6456587609813249536 are 0.139, 0.155, 
         0.258, and 0.195, respectively; significantly below the typical values ($>0.4$) for the sources in our primary sample.
%
%-------------------------------------------------------------------------------------------------------------------------------------------
%
     \begin{table*}
        \centering
         \fontsize{8}{11pt}\selectfont
         \tabcolsep 0.25truecm
         \caption{Processing pipeline validation. Results from \citet{2018ApJ...868...25B}, BR18, and \citet{2018MNRAS.tmp.2466M},  
                  MA19, as well as ours, OURS, for four hypervelocity star candidates, which are not part of our primary sample, are 
                  presented here.}
         \begin{tabular}{ccccccc}
          \hline\hline
            {\it Gaia} DR2 designation & $\alpha$        & $\delta$          & $d_{\rm GC}$               & $v_{\rm GC}$        & $\theta$             
                                       & Source \\
                                       &  (\degr)        &  (\degr)          &        (kpc)               & (km~s$^{-1}$)       &  (\degr)             
                                       &        \\
          \hline
            4326973843264734208        & 248.89229520478 & $-$14.51843538040 & 3.8$\pm$0.4                & 730$\pm$159         & 91.8$\pm$5.3         
                                       & BR18   \\
                                       &                 &                   & 3.842$^{+0.450}_{-0.465}$  & 766$^{+163}_{-122}$ & $-$                  
                                       & MA19   \\
                                       &                 &                   & 3.969$^{+0.464}_{-0.521}$  & 729$^{+168}_{-121}$ & 91.4$^{+5.6}_{-4.6}$ 
                                       & OURS   \\
          \hline
            4395399303719163904        & 258.75009133020 & +08.73144731293   & 8.194$^{+2.309}_{-1.620}$  & 671$^{+136}_{-106}$ & $-$ 
                                       & MA19   \\
                                       &                 &                   & 8.887$^{+4.126}_{-2.105}$  & 714$^{+233}_{-134}$ & 51.5$^{+12.6}_{-13.4}$ 
                                       & OURS   \\
          \hline
            1383279090527227264        & 240.33734815618 & +41.16677411760   & 10.0$\pm$0.9               & 924$\pm$168         & 86.7$\pm$0.6         
                                       & BR18   \\
                                       &                 &                   & 10.064$^{+0.908}_{-0.561}$ & 921$^{+179}_{-124}$ & $-$                  
                                       & MA19   \\
                                       &                 &                   & 10.061$^{+0.903}_{-0.611}$ & 925$^{+178}_{-135}$ & 86.7$^{+0.6}_{-0.6}$ 
                                       & OURS   \\
          \hline
            6456587609813249536        & 317.36089182588 & $-$57.91240021080 & 7.3$\pm$1.7               & 889$\pm$250         & 51.3$\pm$9.0         
                                       & BR18   \\
                                       &                 &                   & 7.222$^{+1.350}_{-0.761}$ & 875$^{+212}_{-155}$ & $-$                  
                                       & MA19   \\
                                       &                 &                   & 7.289$^{+1.676}_{-0.863}$ & 890$^{+256}_{-174}$ & 50.2$^{+9.7}_{-9.1}$ 
                                       & OURS   \\
          \hline
         \end{tabular}
         \label{valid}
     \end{table*}
%
%-------------------------------------------------------------------------------------------------------------------------------------------
%

   \section{Full sample: results\label{Results}}
\begin{figure}
        \centering
         \includegraphics[width=\linewidth]{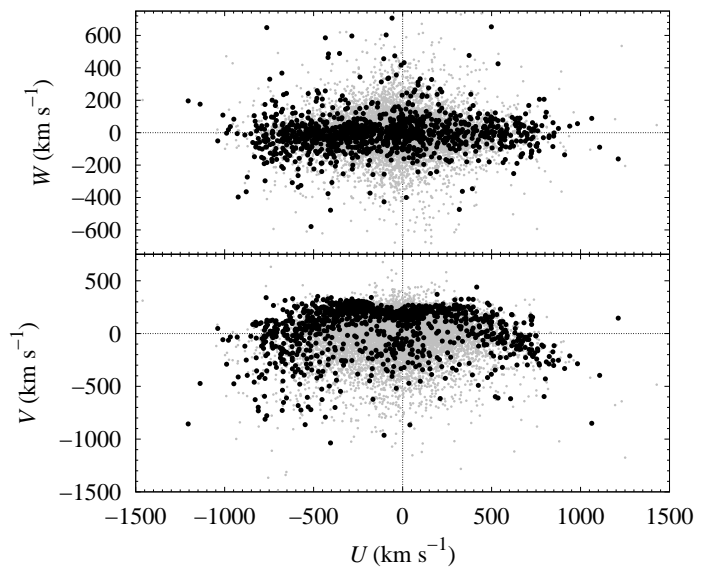}
         \caption{Galactocentric velocity components of the {\it Gaia} DR2 sample of 15\,681 sources (in light grey). Sources with a ratio
                  between the value of the parallax and its uncertainty $>1$ (1150) are plotted in black. The clustering centred at 
                  $(U, V)\sim(-50, -200)$~km~s$^{-1}$ corresponds to the Large Magellanic Cloud ---$(U, V, W)\sim(-58, -221, 209)$~km~s$^{-1}$--- 
                  (mostly) and the Small Magellanic Cloud ---$(U, V, W)\sim(26, -178, 175)$~km~s$^{-1}$ (values corrected for the Solar 
                  motion and the local standard of rest from those in \citealt{2018A&A...616A..12G}).
                 }
         \label{full}
      \end{figure}

      Figure~\ref{full} shows the Galactocentric velocity components of the {\it Gaia} DR2 sample of 15\,681 sources, our primary full 
      sample. The components $U$, $V$, and $W$ are positive in the directions of the Galactic centre, Galactic rotation, and the NGP. The 
      sources in grey show good symmetry in the $U-W$ plane but an obvious asymmetry in the $U-V$ plane. Although the fraction of stars with
      $V<0$ is 33.3\%, their dispersion is wider than that of sources with $V>0$ (meaning that there are more stars with extreme negative 
      values of $V$), which we interpret as a signature of an unbound population as they move retrograde in $V$, this together with the 
      large values of the velocity components. It is however worth to consider here that \citet{2011MNRAS.415.3807S} have shown in their 
      Sect.~3.1 that, when the distance is uncertain, retrograde $V$ components tend to be significantly larger than their prograde 
      counterparts; in other words, the increased dispersion observed for $V<0$ could be a side effect of the large uncertainties in $d$.
      On the other hand and in consistency with our initial hypothesis, 10\,228 sources ($\sim$65\% of our primary full sample) have median 
      Galactocentric speeds above 500~km~s$^{-1}$, which is our threshold for hypervelocity candidacy (in other words, most distant stars are 
      hypervelocity star candidates). If, instead of using the Monte Carlo methodology described above, the usual expressions are applied, 
      13\,293 sources ($\sim$85\% of our primary full sample) have nominal Galactocentric speeds above 500~km~s$^{-1}$. 

      Unfortunately, the uncertainties associated with most sources are very large. Out of the primary full sample, we singled out 1150 
      sources ($\sim$7\%) for which the uncertainty in the value of the parallax is smaller than its nominal value in {\it Gaia} DR2. These 
      sources are plotted in black in Fig.~\ref{full} and they exhibit a pattern of asymmetry consistent with the one found for the primary 
      full sample. Figure~\ref{full}, bottom panel, shows 745 sources with positive $V$-component ($\sim$65\%, these may have an origin in 
      the Milky Way and be part of the halo) and 405 with negative values of $V$; among these, we observe sources that may be in the 
      Magellanic Clouds ---the clustering centred at $(U, V)\sim(-50, -200)$~km~s$^{-1}$. The fact that the sources moving retrograde have 
      higher net speeds suggests that they are indeed intergalactic stars although some substructure (but see the cautionary note above), 
      other than that probably associated with the Magellanic Clouds, is also present in the form of kinematically coherent groups, perhaps 
      debris from disrupted galaxies or actual dwarf galaxies in the process of falling towards the Milky Way galaxy or returning towards
      intergalactic space after experiencing a hyperbolic encounter with the Galaxy.
%
%-------------------------------------------------------------------------------------------------------------------------------------------
%
%
%-------------------------------------------------------------------------------------------------------------------------------------------
%

\begin{figure}
        \centering
         \includegraphics[width=\linewidth]{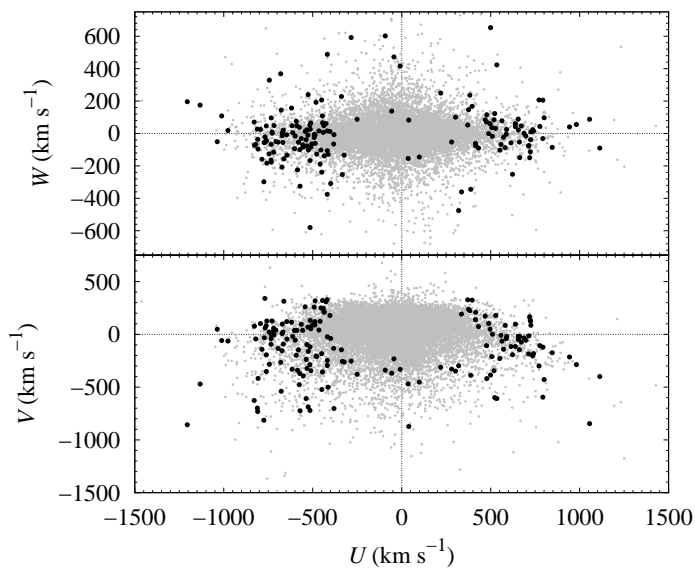}
         \caption{Same as Fig.~\ref{full} but focussing on the best sample including the 174 stars with median Galactocentric distance 
                  greater than 30~kpc and median Galactocentric Galactic velocity greater than 500~km~s$^{-1}$ that verify the criteria for
                  reliable astrometry discussed in Sect.~\ref{SeCr} and shown in Fig.~\ref{quality}. About 64\% of stars in the best sample 
                  have negative median values of the $V$-component.
                 }
         \label{good}
      \end{figure}

      Figure~\ref{good} focusses on the best sample including the 174 stars, sources in black, with median Galactocentric distance greater 
      than 30~kpc and median Galactocentric Galactic velocity greater than 500~km~s$^{-1}$ that verify the criteria for reliable astrometry 
      discussed in Sect.~\ref{SeCr} and shown in Fig.~\ref{quality}. Out of 393 sources with reliable astrometric determination, 168 have 
      $V<0$; out of the 174 stars in the best sample (which are a subset of the 393), 112 have $V<0$. Therefore, most stars in the best 
      sample of 174 follow retrograde trajectories; the percentage of distant hypervelocity candidate stars among the astrometrically 
      well-behaved subsample with $V<0$ is 66.7\%, which is the exact opposite of the trend found for the sample of 1150 sources for which 
      the uncertainty in the value of the parallax is smaller than its nominal value in {\it Gaia} DR2 (Fig.~\ref{full}, bottom panel).
%
%-------------------------------------------------------------------------------------------------------------------------------------------
%
%
%-------------------------------------------------------------------------------------------------------------------------------------------
%

      Figures~\ref{full} and \ref{good}, sources in black, suggest that there is indeed a population of fast and faraway stars that may have 
      an extragalactic origin (mostly negative values of $V$ for the best sample). This is consistent with the discussion in 
      \citet{2016A&A...588A..41C} and the results in \citet{2018MNRAS.tmp.2466M}, and it appears to confirm the presence of a 
      significant population of high-velocity stars of intergalactic provenance. The size of this unbound population must be very large 
      because only the brightest (and therefore most massive) stars are present in the {\it Gaia} DR2 sample of very distant stars. 
%
%-------------------------------------------------------------------------------------------------------------------------------------------
%
      \begin{figure}
        \centering
         \includegraphics[width=\linewidth]{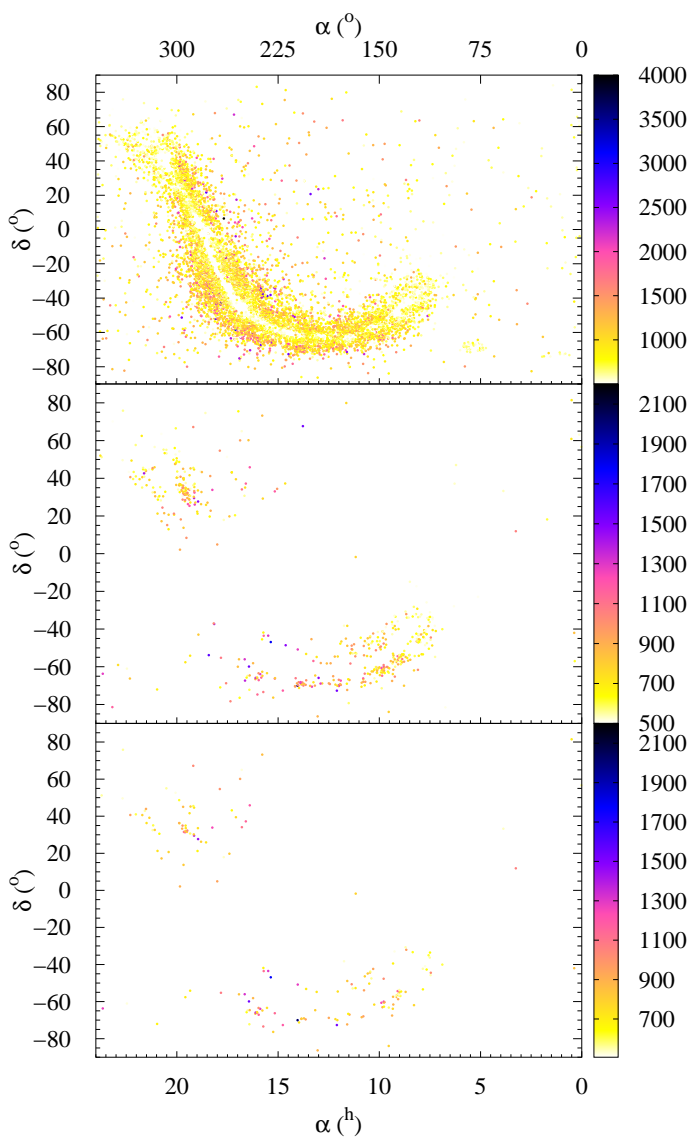}
         \caption{Sources in equatorial coordinates. Distribution in equatorial coordinates of our primary sample of high-velocity and 
                  hypervelocity star candidates (top panel, only the sources with median Galactocentric Galactic velocity greater than 
                  500~km~s$^{-1}$ are shown), those with a ratio between the value of the parallax and its uncertainty $>1$ (middle panel), 
                  and our best sample with most reliable astrometric solution (bottom panel). The colour map shows the values of the 
                  Galactocentric Galactic velocity in km~s$^{-1}$.
                 }
         \label{eq}
      \end{figure}
%
%-------------------------------------------------------------------------------------------------------------------------------------------
%

\begin{figure}
        \centering
         \includegraphics[width=\linewidth]{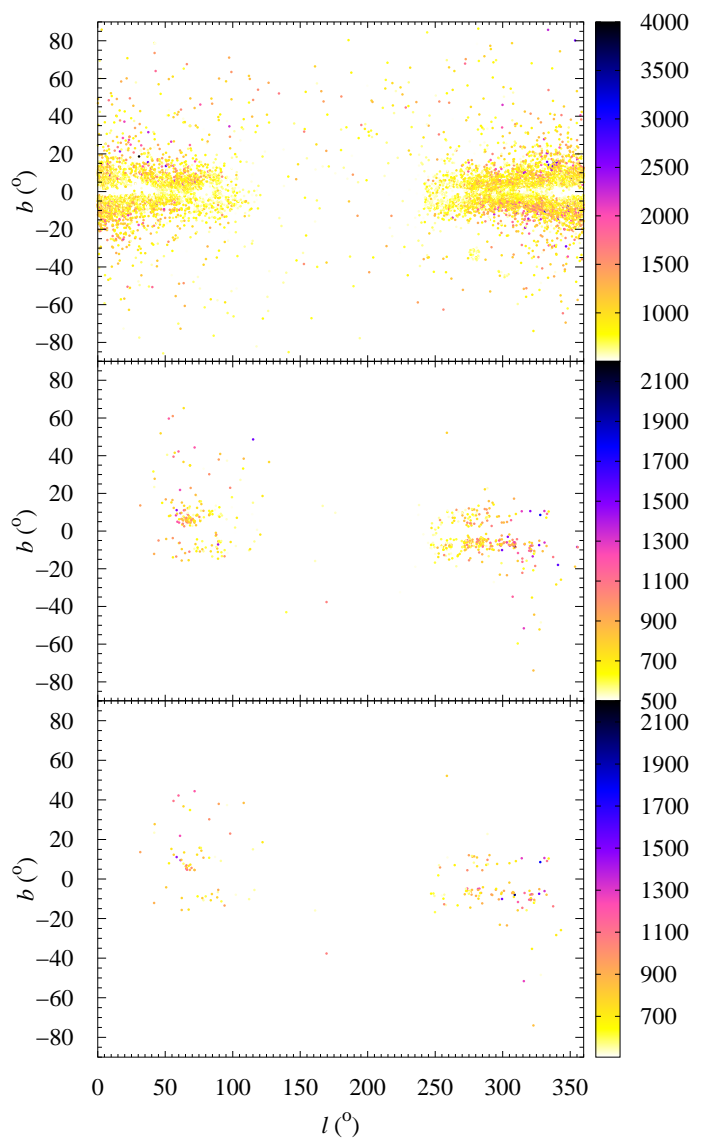}
         \caption{Same as Fig.~\ref{eq} but in Galactic coordinates.
                 }
         \label{gal}
      \end{figure}

      As for the spatial distribution of these distant hypervelocity star candidates, Figs.~\ref{eq} and \ref{gal}, top panels, show the 
      location in the sky of our primary sample; the smaller samples, with lower uncertainties, are displayed in the panels at the bottom. 
      The top panels show that the sources outline the Galactic disc, with most candidates observed projected towards the Galactic bulge 
      (Galactic longitude, $l<50\degr$ and $l>310\degr$ in Fig.~\ref{gal}). There is however a non-negligible fraction of candidates faraway 
      from the disc. \citet{2015ARA&A..53...15B} points out that the previously known hypervelocity stars exhibit statistically 
      significant clustering towards the equatorial coordinates, $\alpha=11^{\rm h}~30^{\rm m}$ and $\delta=3\degr$. Such a concentration is 
      not observed in our full sample and this could be a systematic effect or that the previously known hypervelocity stars belong to a 
      separate population. 
%
%-------------------------------------------------------------------------------------------------------------------------------------------
%
%
%-------------------------------------------------------------------------------------------------------------------------------------------
%

      What is observed in Fig.~\ref{gal} is a scarcity of candidates towards the Galactic anticentre, $l\sim180\degr$, but this must be the 
      result of a selection effect. Ejections from the region of the Galactic centre and bulge moving towards the anticentre will have large 
      values of the radial velocity and comparatively negligible proper motions (many such sources may not have data in {\it Gaia} DR2). On 
      the other hand, runaway exo-Galactic stars observed towards the Galactic anticentre may have virtually zero radial velocity if their 
      paths do not cross that of the Galaxy; in other words, a value may be absent from {\it Gaia} DR2 and therefore they will not be part 
      of our sample due to our selection criteria. The low number of candidates observed towards the Galactic anticentre could be a 
      systematic effect. The presence of clustering is very obvious in the bottom panels of Figs.~\ref{eq} and \ref{gal} and, without a 
      doubt, this must be due to a systematic effect ---some is expected towards the ecliptic poles, see the discussion in 
      \citet{2018A&A...616A...2L}, and the observed locations match approximately. The quality of the astrometric data acquired by the 
      spacecraft for distant sources must be higher towards those directions for technical reasons. It cannot be due to the motion of the 
      Milky Way galaxy ---towards $\alpha=10^{\rm h}~30^{\rm m}$ and $\delta=-24\degr$, with respect to the cosmic microwave background 
      \citep{1993ApJ...419....1K}--- as the Galactic apex (or antapex) is well separated from the concentrations of sources (see 
      Fig.~\ref{eq}, top panel). The fact that most of the candidates appear projected towards the disc suggests that some may have an 
      origin in it as discussed by \citet{2018A&A...620A..48I} and \citet{2018MNRAS.tmp.2466M}. A number of sources appear projected 
      towards the Large Magellanic Cloud ($\alpha=5^{\rm h}~23^{\rm m}$, $\delta=-69\degr~45'$, see Fig.~\ref{eq}, top panel) and they may 
      have an origin in it, like HVS~3 \citep{2018A&A...620A..48I,2019MNRAS.483.2007E} and some B-type hypervelocity stars 
      \citep{2017MNRAS.469.2151B}; these sources are absent from the panels at the bottom due to the application of the various quality 
      criteria discussed in Sect.~\ref{SeCr}. 

   \section{Best samples: results\label{BestRes}}
      The number of high-velocity candidates, $>500$~km~s$^{-1}$, in the sample including 1150 sources (for which the uncertainty in the 
      value of the parallax is smaller than its nominal value) is 539 ($\sim$47\%); for the sample including 393 sources (those with the 
      most reliable astrometric solutions) the number of high-velocity candidates is 174 ($\sim$45\%). In other words, for the sources with 
      the lowest uncertainties the probability of being part of the high-velocity tail associated with the stellar halo or having a true 
      hypervelocity nature is nearly the same. 

\begin{figure*}
        \centering
         \includegraphics[width=0.49\linewidth]{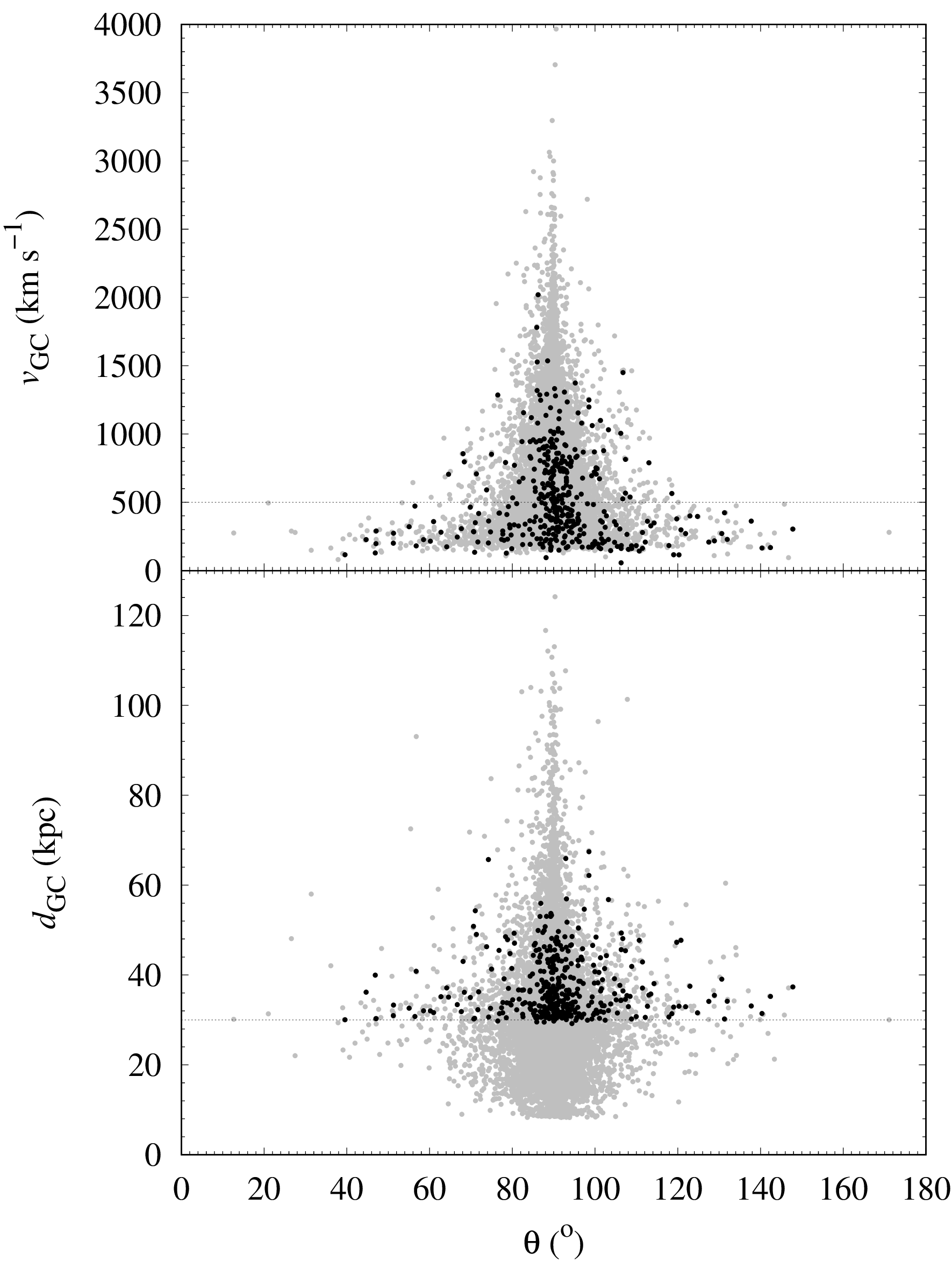}
         \includegraphics[width=0.49\linewidth]{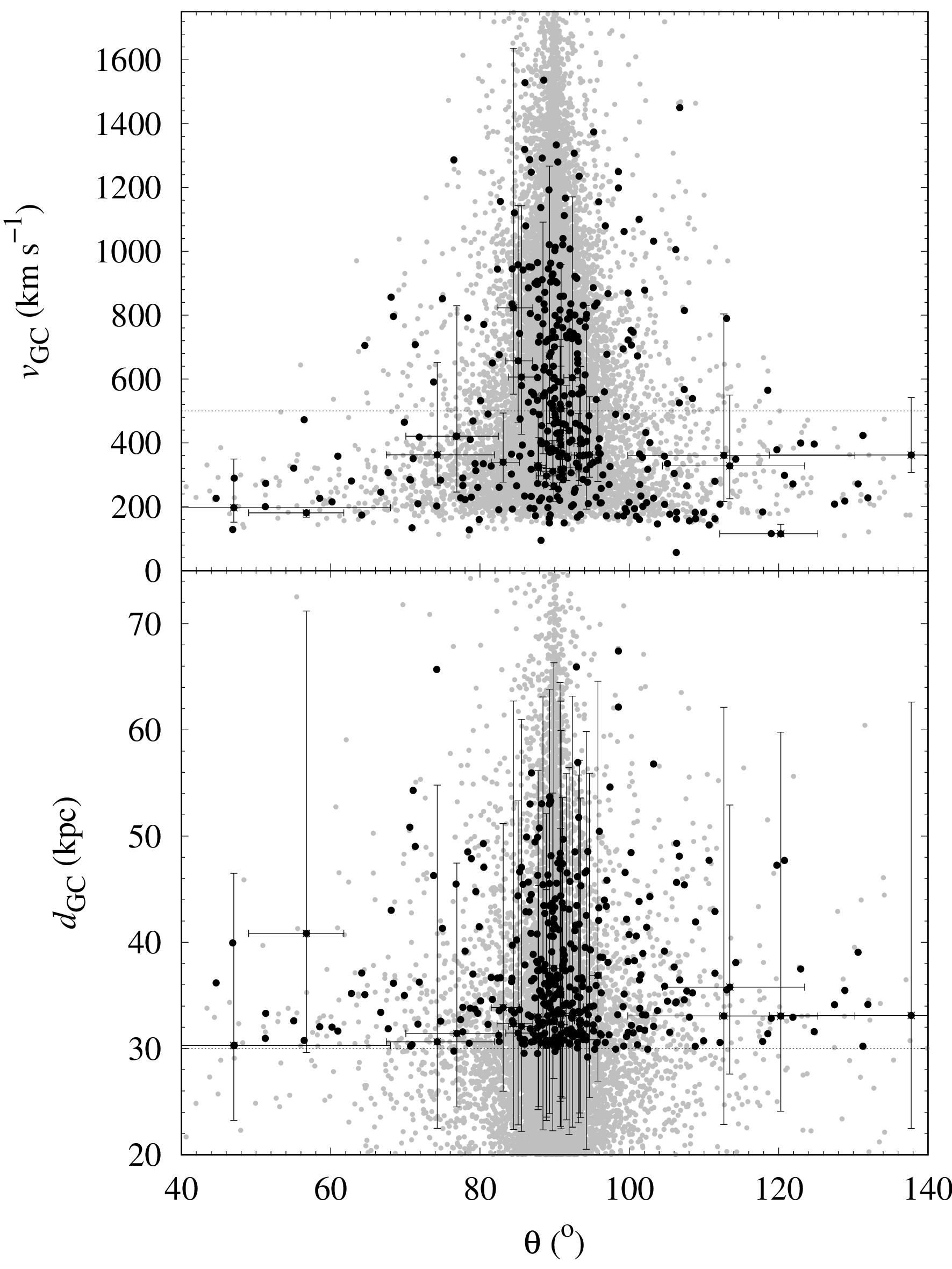}
         \caption{Distances, velocities, and angles. Distribution of Galactocentric distances and velocities as a function of the angle 
                  between the Galactocentric position and velocity vectors for the full sample (in grey) of 15\,681 sources and the one of 
                  393 sources (in black) with the best astrometric solutions (left-hand side panels). The magnified version of the same data 
                  plot (right-hand side panels) includes the error bars for those sources with ratio between the value of the parallax and 
                  its uncertainty $>2$. Our lower cut-off limits, 500~km~s$^{-1}$ and 30~kpc, are indicated by a discontinuous line on each 
                  panel.
                 }
         \label{angle}
      \end{figure*}

      Regarding the origin of the hypervelocity star candidates, Fig.~\ref{angle} shows the distribution of Galactocentric distances and 
      velocities as a function of the angle between the Galactocentric position and velocity vectors. Out of 539 sources, 293 have an angle 
      $<90\degr$; out of 174, the equivalent number is 86. In addition, the mean value is 90\degr, the standard deviation is 6\degr, the 
      median value is 90\degr, and the IQR is about 5\degr for the 539 sources (for the sample of 174, the respective results are 91\degr, 
      7\degr, 90\degr, and about 5\degr). These values suggest that the angular distribution is Gaussian. In any case, the most probable 
      value of the angle between two randomly generated vectors is 90\degr. Nearly half of the best sources may have been ejected from the 
      disc and the others may come from intergalactic space. 
%
%-------------------------------------------------------------------------------------------------------------------------------------------
%
%
%-------------------------------------------------------------------------------------------------------------------------------------------
%

\begin{figure}
        \centering
         \includegraphics[width=\linewidth]{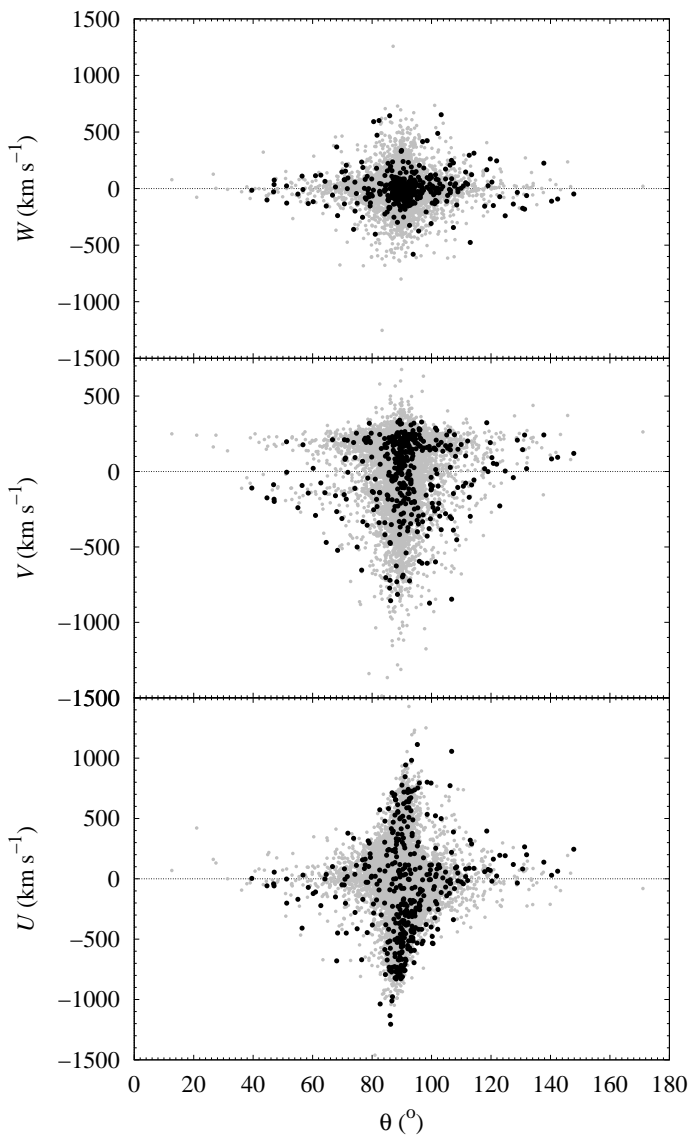}
         \caption{Velocity components and angles. Distribution of Galactocentric Galactic velocity components as a function of the angle 
                  between the Galactocentric position and velocity vectors for the full sample (in grey) of 15\,681 sources and the one of 
                  393 sources (in black) with the best astrometric solutions.  
                 }
         \label{UVWangle}
      \end{figure}

      As for the Galactocentric speed, the average value is 786~km~s$^{-1}$ with a standard deviation of 231~km~s$^{-1}$, a median value of 
      737~km~s$^{-1}$, and an IQR of 308~km~s$^{-1}$ for the 539 sources (for the sample of 174, the respective results are 819~km~s$^{-1}$, 
      246~km~s$^{-1}$, 785~km~s$^{-1}$, and 301~km~s$^{-1}$). Although the overall uncertainties are large even for the sources with the 
      best data (see the error bars in Fig.~\ref{angle}, right-hand panels), the statistical trends are clear and the 539 (174) sources can 
      be considered hypervelocity star candidates although some contamination from the foreground stellar populations is expected. 
      Figure~\ref{angle} also shows that neither runaways from the Galactic centre nor stars moving towards it are present in the sample as 
      these may be comparatively rare (but some high speed, under 500~km~s$^{-1}$, candidates are present in Figs.~\ref{angle} and 
      \ref{UVWangle}). Figure~\ref{UVWangle} shows that a number of stars with $V$$\sim$250~km~s$^{-1}$ have $U$$\sim-$750~km~s$^{-1}$ and
      $\theta$$\sim$90\degr, which again hints at ejections from the Galactic disc.
%
%-------------------------------------------------------------------------------------------------------------------------------------------
%
%
%-------------------------------------------------------------------------------------------------------------------------------------------
%

      We found that 46 sources have uncertainties that are smaller than half the value of the distance, and 10 of them (see 
      Table~\ref{best1}, top section) have values of the Galactocentric speed $>500$~km~s$^{-1}$ (see Table~\ref{best2}, top section). 
      Several of these sources appear to travel together, {\it Gaia} DR2 5303914835370054656, 5304019117178317312, 5317229302706969088, and 
      5317776481532378240, which is somewhat consistent with recent analyses by \citet{2018MNRAS.481.1028H} and 
      \citet{2019arXiv190402159K}. The results for the 10 sources with the lowest uncertainties are still far from satisfactory in terms 
      of errors. The uncertainty in distance (upper limit, meaning 84th percentile) is still close to its median value and the one in 
      velocity is at least 75\% its median value. The bottom section of Tables~\ref{best1} and \ref{best2} shows three relevant examples 
      with the highest values of the Galactocentric velocity and the lowest relative errors, which are at least 80\%. All the stars in 
      Table~\ref{best2} have angles consistent with an origin in the disc (mostly) or outside the Milky Way galaxy. Table~\ref{best1} shows 
      that all the candidates have $|b|<23\degr$, which is suggestive of an origin in the disc or, alternatively, a signal of possible 
      contamination by disc stars with erroneous parallax determinations. The presence of a fraction of false positives among our candidates 
      cannot be discarded (see an obvious case in Sect.~\ref{Discus}), but any star leaving the disc tangent to its original bound 
      trajectory may have an initially low value of its Galactic latitude (see also Figs.~\ref{angle} and \ref{UVWangle}). For example, 
      LAMOST-HVS4 (a hypervelocity star with a likely origin in the Galactic disc) has $b=-17.92\degr$ \citep{2018AJ....156...87L}.
%
%-------------------------------------------------------------------------------------------------------------------------------------------
%
     \begin{table*}
        \centering
         \fontsize{8}{11pt}\selectfont
         \tabcolsep 0.05truecm
         \caption{Hypervelocity star candidates (I). Coordinates, parallaxes, proper motions, radial velocities, and their uncertainties 
                  from {\it Gaia} DR2. The candidates here have uncertainties in parallax that are smaller than half the value of the 
                  parallax (first 10).}
         \begin{tabular}{ccccccccc}
          \hline
          \hline
            {\it Gaia} DR2 designation & $\alpha$        & $\delta$          &    $l$          &    $b$            &  $\pi$              
                                       & $\mu_{\alpha}$     &  $\mu_{\delta}$    & $V_r$              \\
                                       &  (\degr)        &  (\degr)          &  (\degr)        &  (\degr)          &  (mas) 
                                       & (mas yr$^{-1}$)    & (mas yr$^{-1}$)    & (km s$^{-1}$)      \\
          \hline
            1867261981412622976        & 317.50283948290 &   +35.06321358265 &  79.98394750766 &  $-$8.75202945162 & 0.0307$\pm$0.0146
                                       & $-$2.120$\pm$0.020 & $-$3.357$\pm$0.023 &  $-$46.07$\pm$1.47 \\
            2076234920873478272        & 294.95021989885 &   +39.26741709907 &  73.02211904083 &    +8.29595299575 & 0.0308$\pm$0.0146
                                       & $-$2.328$\pm$0.024 & $-$4.329$\pm$0.027 &  $-$73.00$\pm$1.08 \\ 
            2130229806599268480        & 287.78291416193 &   +45.49653404425 &  76.40831708998 &   +15.76370451034 & 0.0295$\pm$0.0141 
                                       & $-$1.842$\pm$0.026 & $-$3.525$\pm$0.029 & $-$115.79$\pm$1.72 \\ 
            5252768161477321472        & 151.81209119622 & $-$63.07207249907 & 285.53624820401 &  $-$5.88769893896 & 0.0319$\pm$0.0152
                                       & $-$4.431$\pm$0.027 &   +3.169$\pm$0.025 &    +64.84$\pm$0.82 \\
            5303914835370054656        & 135.88008255316 & $-$57.43383739232 & 275.88018014633 &  $-$7.07867974501 & 0.0312$\pm$0.0148 
                                       & $-$2.298$\pm$0.031 &   +2.572$\pm$0.033 &   +108.09$\pm$1.37 \\ 
            5304019117178317312        & 137.68599081936 & $-$56.73618453511 & 276.01906096699 &  $-$5.88263899848 & 0.0310$\pm$0.0154
                                       & $-$2.551$\pm$0.031 &   +2.712$\pm$0.034 &   +122.46$\pm$2.91 \\ 
            5317229302706969088        & 133.82399802206 & $-$54.56011892711 & 272.93005017631 &  $-$6.08394727436 & 0.0296$\pm$0.0142
                                       & $-$2.743$\pm$0.027 &   +3.231$\pm$0.025 &    +93.75$\pm$1.13 \\ 
            5317776481532378240        & 130.49490546902 & $-$54.79662905248 & 271.90741854172 &  $-$7.73602796070 & 0.0317$\pm$0.0158
                                       & $-$3.186$\pm$0.031 &   +4.565$\pm$0.028 &   +114.04$\pm$0.33 \\ 
            5489467833544327040        & 117.63555537419 & $-$53.40035704153 & 266.59793639347 & $-$13.34069255350 & 0.0335$\pm$0.0145
                                       & $-$0.908$\pm$0.030 &   +4.537$\pm$0.028 &   +129.16$\pm$0.95 \\ 
            5531793499296526336        & 116.33885534981 & $-$44.89073152454 & 258.52760575222 & $-$10.06954996387 & 0.0354$\pm$0.0174
                                       & $-$1.462$\pm$0.031 &   +3.101$\pm$0.029 &    +80.86$\pm$3.35 \\ 
          \hline
          \hline
            2040146269183375232        & 284.24996846082 &   +27.66842104956 &  58.47199343295 &   +11.10518803277 & 0.0248$\pm$0.0146 
                                       & $-$4.244$\pm$0.020 & $-$7.082$\pm$0.026 & $-$364.65$\pm$0.96 \\ 
            2255126837089768192        & 287.82340994615 &   +67.16711968401 &  98.10331412498 &   +22.97752933362 & 0.0181$\pm$0.0126
                                       & $-$4.060$\pm$0.026 & $-$0.627$\pm$0.026 & $-$388.70$\pm$0.79 \\ 
            5846959127210423424        & 210.01295570953 & $-$69.82803148077 & 308.82316018688 &  $-$7.74494766675 & 0.0274$\pm$0.0190
                                       & $-$6.796$\pm$0.027 & $-$0.722$\pm$0.029 &     +8.50$\pm$1.05 \\ 
          \hline
         \end{tabular}
         \label{best1}
     \end{table*}
%
%-------------------------------------------------------------------------------------------------------------------------------------------
%
%
%-------------------------------------------------------------------------------------------------------------------------------------------
%
     \begin{table}
        \centering
         \fontsize{8}{12pt}\selectfont
         \tabcolsep 0.15truecm
         \caption{Hypervelocity star candidates (II). Galactocentric distance and velocity, and the angle between the position and velocity 
                  vectors (median values and uncertainties derived from the 16th and 84th percentiles) have been computed using the Monte 
                  Carlo approach described in the text. RUWE index from \url{http://gaia.ari.uni-heidelberg.de/singlesource.html}.}
         \begin{tabular}{ccccc}
          \hline
          \hline
            {\it Gaia} DR2 designation & $d_{\rm GC}$           & $v_{\rm GC}$          & $\theta$                & RUWE  \\
                                       & (kpc)                  & (km~s$^{-1}$)         &  (\degr)                &       \\
          \hline
            1867261981412622976        & 32.2$_{-10.0}^{+28.8}$ &  608$_{-181}^{+533}$  &  85.6$_{-1.7}^{+2.0}$   & 0.944 \\
            2076234920873478272        & 30.9$_{-9.8}^{+28.1}$  &  711$_{-229}^{+657}$  &  91.1$_{-0.5}^{+0.5}$   & 0.973 \\
            2130229806599268480        & 33.1$_{-10.5}^{+30.4}$ &  603$_{-199}^{+573}$  &  92.4$_{-1.1}^{+1.0}$   & 0.964 \\
            5252768161477321472        & 30.2$_{-9.5}^{+28.2}$  &  751$_{-249}^{+732}$  &  87.9$_{-0.6}^{+0.9}$   & 1.048 \\ 
            5303914835370054656        & 32.2$_{-9.7}^{+27.8}$  &  506$_{-158}^{+456}$  &  90.9$_{-0.6}^{+0.9}$   & 0.967 \\
            5304019117178317312        & 32.4$_{-10.2}^{+31.1}$ &  545$_{-182}^{+553}$  &  88.4$_{-0.5}^{+0.6}$   & 0.999 \\ 
            5317229302706969088        & 34.3$_{-10.5}^{+30.3}$ &  672$_{-212}^{+611}$  &  89.3$_{-0.2}^{+0.2}$   & 0.886 \\
            5317776481532378240        & 32.3$_{-9.9}^{+30.3}$  &  821$_{-270}^{+810}$  &  84.5$_{-2.2}^{+2.6}$   & 0.942 \\
            5489467833544327040        & 31.4$_{-8.6}^{+22.1}$  &  655$_{-193}^{+491}$  &  85.1$_{-1.7}^{+2.0}$   & 0.927 \\
            5531793499296526336        & 31.0$_{-8.8}^{+26.4}$  &  523$_{-141}^{+426}$  &  92.8$_{-1.2}^{+0.9}$   & 0.941 \\
          \hline
          \hline
            2040146269183375232        & 36.7$_{-14.4}^{+52.8}$ & 1459$_{-573}^{+2073}$ & 106.6$_{-10.1}^{+11.0}$ & 0.984 \\ 
            2255126837089768192        & 56.8$_{-22.2}^{+92.6}$ & 1032$_{-429}^{+1827}$ & 103.2$_{-9.3}^{+9.8}$   & 0.939 \\
            5846959127210423424        & 31.8$_{-14.2}^{+62.8}$ & 1024$_{-464}^{+2040}$ &  88.7$_{-0.9}^{+0.9}$   & 0.952 \\
          \hline
         \end{tabular}
         \label{best2}
     \end{table}
%
%-------------------------------------------------------------------------------------------------------------------------------------------
%
     
      The vast majority of the candidates in Table~\ref{best1} do not pass the astrometric quality filters discussed in Sect.~\ref{SeCr}. If 
      we focus on the 174 sources, which have the best astrometric solutions, we obtain the best 10 candidates in Tables~\ref{vbest1} and 
      \ref{vbest2}. Only one star, {\it Gaia}~DR2 2130229806599268480, is also included in Tables~\ref{best1} and \ref{best2}. This shows 
      that, for our data, having a well-behaved astrometric solution and $\sigma_{\pi}/\pi<0.5$ are seldom compatible. The sample in 
      Tables~\ref{vbest1} and \ref{vbest2} includes one candidate with $|b|>30\degr$, {\it Gaia} DR2 1696697285206197248, which may have an
      origin in the Galactic bulge region ($\theta$$\sim$68\degr). In addition, {\it Gaia} DR2 5348384273914768000 emerges as a clear 
      candidate to have an exo-Galactic origin ($\theta$$\sim$119\degr).  
%
%-------------------------------------------------------------------------------------------------------------------------------------------
%
     \begin{table*}
        \centering
         \fontsize{8}{11pt}\selectfont
         \tabcolsep 0.05truecm
         \caption{Best hypervelocity star candidates (I). The candidates here belong to the sample of 174 sources, which are probably 
                  distant and have the best astrometric solutions. Only one star, {\it Gaia}~DR2 2130229806599268480, is also included in
                  Tables~\ref{best1} and \ref{best2}. Data source as in Table~\ref{best1}.}
         \begin{tabular}{ccccccccc}
          \hline
          \hline
            {\it Gaia} DR2 designation & $\alpha$        & $\delta$          &    $l$          &    $b$            &  $\pi$              
                                       & $\mu_{\alpha}$     &  $\mu_{\delta}$    & $V_r$              \\
                                       &  (\degr)        &  (\degr)          &  (\degr)        &  (\degr)          &  (mas) 
                                       & (mas yr$^{-1}$)    & (mas yr$^{-1}$)    & (km s$^{-1}$)      \\
          \hline
            1696697285206197248        & 236.71959038696 &   +73.23047076409 & 108.14355244960 &   +38.44334319408 & 0.0248$\pm$0.0134
                                       & $-$4.688$\pm$0.026 &   +1.991$\pm$0.024 &    +11.81$\pm$1.69 \\
            2045060021021275776        & 292.52188374790 &   +31.54338716612 &  65.21699800964 &   +06.38673084189 & 0.0286$\pm$0.0155
                                       & $-$2.671$\pm$0.026 & $-$4.648$\pm$0.027 & $-$104.15$\pm$0.72 \\
            2079869356551980672        & 295.32788216218 &   +45.04160856299 &  78.31989589146 &   +10.77144848514 & 0.0257$\pm$0.0157
                                       & $-$2.290$\pm$0.029 & $-$4.337$\pm$0.027 & $-$207.88$\pm$1.60 \\
            2101802964257385984        & 290.56036760943 &   +41.60838045004 &  73.60284663359 &   +12.33450825973 & 0.0275$\pm$0.0168
                                       & $-$2.318$\pm$0.029 & $-$3.927$\pm$0.030 & $-$111.10$\pm$2.57 \\
            2130229806599268480        & 287.78291416193 &   +45.49653404425 &  76.40831708998 &   +15.76370451034 & 0.0295$\pm$0.0141
                                       & $-$1.842$\pm$0.026 & $-$3.525$\pm$0.029 & $-$115.79$\pm$1.72 \\
            2186552770772130048        & 307.76357356019 &   +55.06179129105 &  91.18635454905 &   +09.15879626453 & 0.0286$\pm$0.0201
                                       & $-$1.964$\pm$0.043 & $-$2.591$\pm$0.038 & $-$153.88$\pm$0.78 \\
            5348384273914768000        & 169.73145311955 & $-$52.90554045217 & 289.10868902638 &   +07.46532297730 & 0.0302$\pm$0.0179
                                       &   +2.842$\pm$0.027 &   +0.033$\pm$0.026 &    +75.05$\pm$0.79 \\
            5365388599188363648        & 155.82929212599 & $-$47.49147237014 & 278.65087406165 &   +08.27589925726 & 0.0276$\pm$0.0173
                                       & $-$2.758$\pm$0.030 &   +2.084$\pm$0.031 &   +120.04$\pm$1.41 \\
            5370493837834558976        & 173.72835355643 & $-$48.87607199491 & 290.19714079581 &   +12.09429343550 & 0.0302$\pm$0.0169
                                       & $-$5.179$\pm$0.025 & $-$2.647$\pm$0.022 &    +34.40$\pm$0.74 \\
            5641968206535885696        & 130.52236743872 & $-$30.73543082547 & 252.84426142858 &   +07.01406877053 & 0.0298$\pm$0.0199
                                       & $-$2.689$\pm$0.028 &   +1.614$\pm$0.030 &   +148.29$\pm$2.40 \\
          \hline
         \end{tabular}
         \label{vbest1}
     \end{table*}
%
%-------------------------------------------------------------------------------------------------------------------------------------------
%
%
%-------------------------------------------------------------------------------------------------------------------------------------------
%
     \begin{table}
        \centering
         \fontsize{8}{12pt}\selectfont
         \tabcolsep 0.15truecm
         \caption{Best hypervelocity star candidates (II). Data source as in Table~\ref{best2}.} 
         \begin{tabular}{ccccc}
          \hline
          \hline
            {\it Gaia} DR2 designation & $d_{\rm GC}$           & $v_{\rm GC}$          & $\theta$                & RUWE  \\
                                       & (kpc)                  & (km~s$^{-1}$)         &  (\degr)                &       \\
          \hline
            1696697285206197248        & 43.0$_{-13.8}^{+44.8}$ &  856$_{-324}^{+1093}$ &  68.1$_{-13.0}^{+12.2}$ & 1.171 \\
            2045060021021275776        & 32.3$_{-11.7}^{+39.2}$ &  802$_{-304}^{+1003}$ &  94.3$_{-2.4}^{+2.3}$   & 0.905 \\
            2079869356551980672        & 38.2$_{-14.2}^{+54.4}$ &  869$_{-342}^{+1269}$ &  99.8$_{-6.1}^{+6.0}$   & 0.968 \\
            2101802964257385984        & 34.9$_{-13.1}^{+50.0}$ &  733$_{-287}^{+1086}$ &  93.0$_{-1.9}^{+1.7}$   & 1.063 \\
            2130229806599268480        & 33.1$_{-10.5}^{+30.1}$ &  604$_{-199}^{+567}$  &  92.3$_{-1.1}^{+1.0}$   & 0.964 \\
            2186552770772130048        & 35.9$_{-13.5}^{+58.1}$ &  559$_{-212}^{+894}$  &  93.7$_{-2.6}^{+2.4}$   & 0.918 \\
            5348384273914768000        & 31.4$_{-11.6}^{+43.9}$ &  564$_{-155}^{+568}$  & 118.6$_{-16.0}^{+13.9}$ & 0.961 \\
            5365388599188363648        & 36.0$_{-13.5}^{+52.2}$ &  565$_{-220}^{+862}$  &  89.4$_{-0.2}^{+0.4}$   & 0.961 \\
            5370493837834558976        & 31.4$_{-11.3}^{+39.4}$ &  831$_{-315}^{+1097}$ &  93.8$_{-2.4}^{+2.8}$   & 1.009 \\
            5641968206535885696        & 36.6$_{-12.7}^{+52.6}$ &  550$_{-190}^{+784}$  &  87.2$_{-1.9}^{+1.9}$   & 1.009 \\
          \hline
         \end{tabular}
         \label{vbest2}
     \end{table}
%
%-------------------------------------------------------------------------------------------------------------------------------------------
%

   \section{Statistical significance\label{StatSig}}
      Data from {\it Gaia} DR2 for sources beyond 30~kpc are affected by significant uncertainties, very large in the case of the parallax
      values (and therefore for the distances), although the values of the radial velocity and those of the components of the proper motion 
      are sufficiently reliable in most cases. Given the large values of the uncertainties in the values of the Galactocentric distance and 
      particularly in those of the velocity, even in the case of the most promising candidates (see Tables~\ref{best1}, \ref{best2}, 
      \ref{vbest1}, and \ref{vbest2}), our results must be interpreted with some caution. However, the fact that Fig.~\ref{full} shows a 
      consistent asymmetry in the values of the $V$ component of the Galactocentric velocity suggests that most of the sources in our 
      samples could be unbound from the Milky Way galaxy. This is a statistically significant result that receives further support from the 
      fact (see Sect.~\ref{FarAway}) that the net effect of the Eddington-Trumpler-Weaver bias makes the stars appear closer than they 
      really are. If most of the sources in Tables~\ref{best1}, \ref{best2}, \ref{vbest1}, and \ref{vbest2} are more distant than estimated, 
      their Galactocentric velocities are also probably higher; in other words, the median values of $d_{\rm GC}$ and $v_{\rm GC}$ in 
      Tables~\ref{best2} and \ref{vbest2} are mostly and probably lower limits and their true values could be close to the 84th percentile 
      as shown in the tables.  

      It can be argued that our criteria to select hypervelocity candidates is perhaps too optimistic and that the statistical significance
      of our best candidates is somewhat marginal, however our approach is in line with the ones used in other studies; for example, one
      recent discovery by LAMOST has a Galactic rest-frame radial velocity of 408~km~s$^{-1}$ at a Galactocentric radius of $\sim$30~kpc
      \citep{2017ApJ...847L...9H}. Several candidates in Tables~\ref{best2}, \ref{vbest2}, and \ref{vbesth2} have comparable values even 
      when the uncertainties are factored in. It is however true that LAMOST candidates are more reliable because their large velocities are 
      virtually unaffected by the uncertainty in the stellar distance of their stars.

   \section{Discussion\label{Discus}}
      Most of the distant sources with Galactocentric speed under 500~km~s$^{-1}$ could be part of the regular halo of the Milky Way or be
      part of an outer spiral arm such as the one described by \citet{2011ApJ...734L..24D}. Star formation away from the Galactic disc is
      possible and it is a well-studied subject (for example, see \citealt{2008ApJ...685L.125D}). Those in the halo may also have an 
      external origin as the Milky Way cannibalizes nearby dwarf galaxies \citep{1994Natur.370..194I}. \citet{2018ApJ...869L..31D} 
      find 24 high-velocity stars using {\it Gaia} DR2 data and spectroscopy from the LAMOST Data Release 5 and five of them appear to 
      have an origin in the tidal debris of a disrupted dwarf galaxy, the rest seem to come from the Galactic disc. LAMOST-HVS1 has an 
      intrinsic ejection velocity of 568$_{-17}^{+19}$~km~s$^{-1}$ and it may come from a yet-to-be-discovered massive young star cluster 
      located near the Norma spiral arm, although it has $b=+35.41\degr$ \citep{2019ApJ...873..116H}. LAMOST-HVS4 could be following a 
      path with an origin in the disc (not the Galactic centre) with an outbound velocity as high as 590$\pm$7~km~s$^{-1}$ 
      \citep{2018AJ....156...87L}. In general, sources with Galactocentric speed above 500~km~s$^{-1}$ cannot be explained within the 
      framework of synthetic models such as \textsc{galaxia} \citep{2011ApJ...730....3S}, they must be the result of scattering or 
      ejection events here in the Milky Way galaxy, within the Local Group, or elsewhere in the Universe. 

      Mechanisms capable of explaining the high-velocity stars coming from the disc include the dynamically driven double-degenerate 
      double-detonation scenario \citep{2018ApJ...865...15S}, rapid mass-loss from a companion in a close binary including a supernova 
      event (for example, see \citealt{2015MNRAS.448L...6T}) and interactions with intermediate-mass black holes within rich star clusters 
      \citep{2015ARA&A..53...15B,2018MNRAS.479.2789B,2018A&A...620A..48I}). The putative population of 
      extragalactic origin may come from similar events outside our galaxy but also have more exotic sources such as disrupted galaxies, 
      encounters with runaway supermassive black holes or intergalactic supernovae. Hypervelocity stars can be easily produced at the 
      centres of galaxies via close encounters between binary stars and supermassive black holes; for example, HV~1 from the centre of the 
      Milky Way \citep{2005ApJ...622L..33B} or HVS~3 from the centre of the Large Magellanic Cloud \citep{2018A&A...620A..48I,
      2019MNRAS.483.2007E}, although a fly-by with a supermassive black hole may occur outside of a galaxy. Runaway supermassive 
      black holes are expected to exist \citep{1972ApL....11...87S,1974ApJ...190..253S,1990ApJ...348..412M,2007ApJ...666L..13B,
      2008ApJ...678..780G} ---but also intermediate-mass ones \citep{2012MNRAS.422..841A}--- and, in their way towards intergalactic space, 
      they may scatter a significant number of stars. In addition, they may trigger star formation themselves on free-floating giant 
      molecular clouds \citep{2008ApJ...677L..47D}. Galaxy-less star formation may produce intergalactic supernovae 
      \citep{2003AJ....125.1087G}, which in turn may produce runaway stars.

      One of the strongest objections that one may make regarding the interpretation of our results is that the vast majority of candidates 
      in Tables~\ref{best1} and \ref{vbest1} have $|b|<23$\degr. Tables~\ref{vbesth1} and \ref{vbesth2} show candidates from the sample of
      174 sources with well-behaved astrometric solutions that have $|b|>30\degr$. One of them, {\it Gaia} DR2 1696697285206197248, was 
      already included in Tables~\ref{vbest1} and \ref{vbest2}. Although in smaller numbers, candidates located well away from the Galactic
      disc are present in our samples as well (see also Fig.~\ref{gal}).
%
%-------------------------------------------------------------------------------------------------------------------------------------------
%
     \begin{table*}
        \centering
         \fontsize{8}{11pt}\selectfont
         \tabcolsep 0.05truecm
         \caption{Best high Galactic latitude ($|b|>30\degr$) hypervelocity star candidates (I). The candidates here belong to the sample of 
                  174 sources, which are probably distant and have the best astrometric solutions. Data source as in Table~\ref{best1}.}
         \begin{tabular}{ccccccccc}
          \hline
          \hline
            {\it Gaia} DR2 designation & $\alpha$        & $\delta$          &    $l$          &    $b$            &  $\pi$
                                       & $\mu_{\alpha}$     &  $\mu_{\delta}$    & $V_r$              \\
                                       &  (\degr)        &  (\degr)          &  (\degr)        &  (\degr)          &  (mas)
                                       & (mas yr$^{-1}$)    & (mas yr$^{-1}$)    & (km s$^{-1}$)      \\
          \hline
            1314483506970781824        & 251.80005602927 &   +34.09044416258 &  56.12871617510 &   +39.48243476824 & 0.0211$\pm$0.0192
                                       & $-$2.876$\pm$0.028 & $-$4.863$\pm$0.037 & $-$250.07$\pm$0.73 \\
            1331376987735140864        & 248.87279142999 &   +37.21051070025 &  59.84507313050 &   +42.21970456491 & 0.0225$\pm$0.0153
                                       & $-$2.472$\pm$0.022 & $-$5.669$\pm$0.026 & $-$116.30$\pm$2.12 \\
            1353387698695201792        & 256.23459332334 &   +39.47807346868 &  63.53398480591 &   +36.70985332927 & 0.0202$\pm$0.0125
                                       & $-$3.721$\pm$0.021 &   +0.654$\pm$0.023 & $-$115.04$\pm$0.89 \\
            1360405705322797312        & 259.28463512742 &   +43.15993767132 &  68.36783292048 &   +34.85872854418 & 0.0275$\pm$0.0211
                                       & $-$4.538$\pm$0.037 & $-$0.371$\pm$0.040 & $-$116.75$\pm$1.02 \\
            1409724578557941248        & 245.96492962581 &   +45.85548579341 &  71.74122097135 &   +44.37334749643 & 0.0154$\pm$0.0137
                                       & $-$1.060$\pm$0.023 & $-$4.180$\pm$0.031 & $-$220.07$\pm$1.98 \\
            1417386319177949824        & 268.14211061898 &   +54.63524890374 &  82.57367537620 &   +30.17952098747 & 0.0290$\pm$0.0218
                                       & $-$3.379$\pm$0.048 & $-$5.631$\pm$0.042 & $-$284.40$\pm$1.59 \\
            1438071495854426880        & 253.13412753466 &   +60.12250785829 &  89.65213628261 &   +37.96715431762 & 0.0272$\pm$0.0246
                                       & $-$6.268$\pm$0.038 & $-$0.483$\pm$0.050 &  $-$96.89$\pm$0.72 \\
            1635042136319192576        & 251.91264273176 &   +64.99736996879 &  95.80257919212 &   +37.35065366127 & 0.0213$\pm$0.0146
                                       & $-$2.434$\pm$0.024 & $-$1.501$\pm$0.030 & $-$312.25$\pm$1.30 \\
            1696697285206197248        & 236.71959038696 &   +73.23047076409 & 108.14355244960 &   +38.44334319408 & 0.0248$\pm$0.0134
                                       & $-$4.688$\pm$0.026 &   +1.991$\pm$0.024 &    +11.81$\pm$1.69 \\
            3791385339777413504        & 167.47408084540 & $-$01.78401513372 & 258.65578792586 &   +52.13596520194 & 0.0294$\pm$0.0230
                                       & $-$0.712$\pm$0.047 & $-$5.779$\pm$0.030 &  $-$90.79$\pm$1.69 \\
          \hline
         \end{tabular}
         \label{vbesth1}
     \end{table*}
%
%-------------------------------------------------------------------------------------------------------------------------------------------
%
%
%-------------------------------------------------------------------------------------------------------------------------------------------
%
     \begin{table}
        \centering
         \fontsize{8}{12pt}\selectfont
         \tabcolsep 0.15truecm
         \caption{Best high Galactic latitude ($|b|>30\degr$) hypervelocity star candidates (II). Data source as in Table~\ref{best2}.}
         \begin{tabular}{ccccc}
          \hline
          \hline
            {\it Gaia} DR2 designation & $d_{\rm GC}$            & $v_{\rm GC}$           & $\theta$                & RUWE  \\
                                       & (kpc)                   & (km~s$^{-1}$)          &  (\degr)                &       \\
          \hline
            1314483506970781824        & 43.8$_{-21.2}^{+97.0}$  & 1100$_{-588}^{+2653}$  & 101.3$_{-10.4}^{+12.9}$ & 0.880 \\
            1331376987735140864        & 42.1$_{-17.4}^{+74.0}$  & 1155$_{-522}^{+2201}$  &  95.9$_{-4.2}^{+4.8}$   & 1.058 \\
            1353387698695201792        & 47.1$_{-18.3}^{+70.2}$  &  771$_{-328}^{+1272}$  &  80.5$_{-7.9}^{+6.3}$   & 1.075 \\
            1360405705322797312        & 34.6$_{-14.9}^{+68.5}$  &  650$_{-330}^{+1527}$  &  81.7$_{-9.6}^{+6.8}$   & 1.004 \\
            1409724578557941248        & 62.1$_{-28.6}^{+133.0}$ & 1198$_{-596}^{+2767}$  &  98.5$_{-7.7}^{+8.6}$   & 0.900 \\
            1417386319177949824        & 34.3$_{-13.8}^{+61.6}$  & 1005$_{-447}^{+1946}$  & 106.2$_{-12.2}^{+12.6}$ & 0.746 \\
            1438071495854426880        & 36.6$_{-15.7}^{+73.0}$  &  944$_{-500}^{+2264}$  &  82.3$_{-7.8}^{+7.0}$   & 0.979 \\
            1635042136319192576        & 48.1$_{-18.4}^{+78.0}$  &  525$_{-242}^{+1095}$  & 106.7$_{-12.2}^{+15.3}$ & 1.041 \\
            1696697285206197248        & 43.0$_{-13.8}^{+44.8}$  &  856$_{-324}^{+1093}$  &  68.1$_{-13.0}^{+12.2}$ & 1.171 \\
            3791385339777413504        & 35.5$_{-13.8}^{+63.1}$  &  789$_{-368}^{+1821}$  & 113.0$_{-18.3}^{+20.8}$ & 0.923 \\
          \hline
         \end{tabular}
         \label{vbesth2}
     \end{table}
%
%-------------------------------------------------------------------------------------------------------------------------------------------
%

\begin{figure*}
        \centering
         \includegraphics[width=0.49\linewidth]{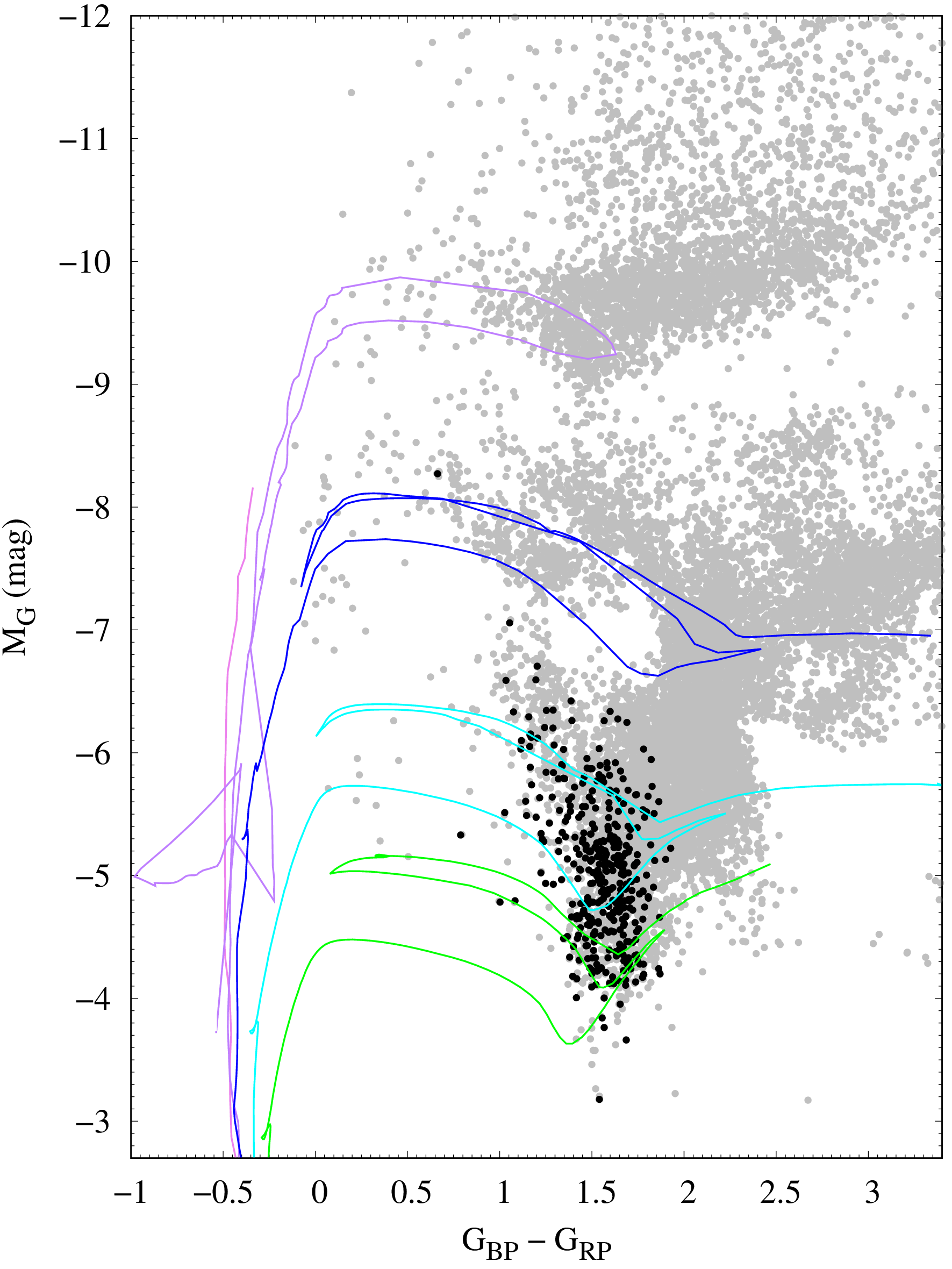}
         \includegraphics[width=0.497\linewidth]{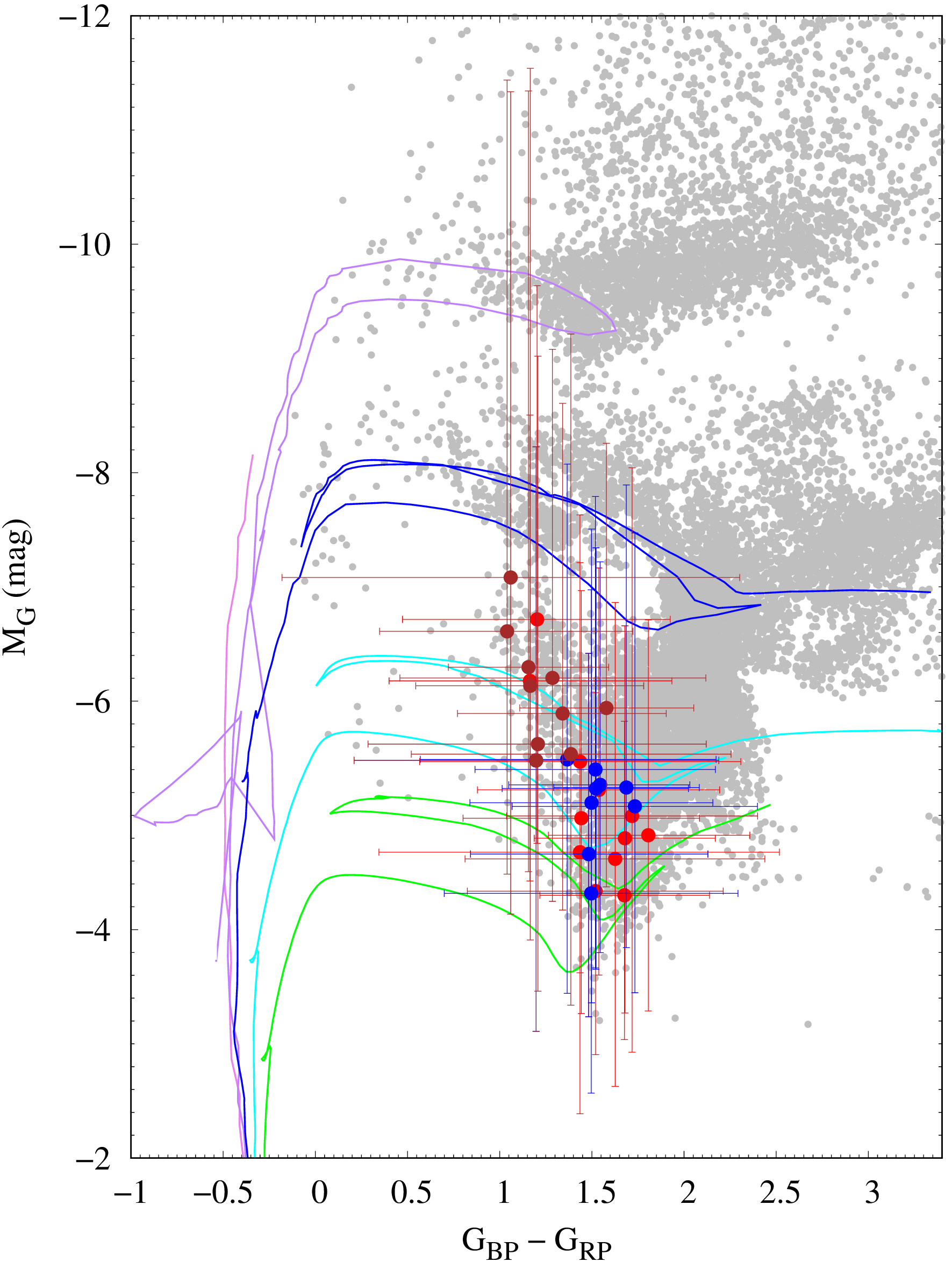}
         \caption{Colours, magnitudes, and theoretical isochrones. Colour-magnitude diagram (CMD) for sources in Fig.~\ref{full} (left-hand 
                  side panel), full sample of 15\,681 sources (in grey) and the one of 393 sources (in black). Five isochrones of ages 1~Myr 
                  (in violet), 5~Myr (in purple), 10~Myr (in blue), 25~Myr (in cyan), and 50~Myr (in green), and solar metallicity are 
                  plotted as a reference (see the text for details). CMD for the candidates in Tables~\ref{best1} (in red), \ref{vbest1} (in 
                  blue), and \ref{vbesth1} (in brown) with error bars (right-hand side panel).
                 }
         \label{cmd}
      \end{figure*}

      On the other hand, our sample is made of sources with estimated values of both the line-of-sight extinction and reddening in {\it Gaia} 
      DR2; it is therefore possible to construct a colour-magnitude diagram (CMD) with the data. Figure~\ref{cmd}, left-hand side panel, 
      shows the resulting CMD: in grey, the full sample, in black, the 393 stars with the best astrometric solutions. This CMD has been 
      obtained as the ones in Fig.~5 of \citet{2018A&A...616A..10G}, Fig.~19 of \citet{2018A&A...616A...8A}, or Fig.~3 in 
      \citet{2018MNRAS.481L..64D}. The median values were computed from the Monte Carlo sampling described above by assuming that the
      standard deviation in the estimated values of the line-of-sight extinction $A_G$ and reddening $E(G_{\rm BP}-G_{\rm RP})$ from 
      {\it Gaia} DR2 can be approximated by the difference between their respective 84th and 16th percentiles (which is a rather pessimistic
      assumption). The same approximation was used to estimate the error bars in Fig.~\ref{cmd}, right-hand side panel. As a reference, 
      five relevant PARSEC v1.2S + COLIBRI S\_35 \citep{2012MNRAS.427..127B,2017ApJ...835...77M,
      2019MNRAS.485.5666P}\footnote{\url{http://stev.oapd.inaf.it/cgi-bin/cmd}} isochrones ---of ages 1~Myr (in violet), 5~Myr (in 
      purple), 10~Myr (in blue), 25~Myr (in cyan), and 50~Myr (in green)--- are also plotted. The entries in Tables~\ref{best1}, \ref{vbest1}, 
      and \ref{vbesth1} (see Fig.~\ref{cmd}, right-hand side panel) appear to be consistent with being massive stars, younger than about 
      50~Myr. This young age together with their high velocities imply that they have been born in the Local Group, probably within 100~kpc 
      from the centre of the Milky Way galaxy (unless they are blue stragglers). The fact that all the best candidates are consistently 
      located within the CMD, in the region of the classical Cepheids, can be considered as a robust supporting argument in favour of the 
      conclusions obtained in Sect.~\ref{FarAway}, where the effect of the Eddington-Trumpler-Weaver bias was discussed. Even sources out of 
      the best sample exhibit conspicuous sequences that outline the various, temporarily stable, phases of stellar evolution for massive 
      stars, including the loci of the luminous blue variables and yellow hypergiants with absolute magnitudes close to $-$10~mag. 
%
%-------------------------------------------------------------------------------------------------------------------------------------------
%
%
%-------------------------------------------------------------------------------------------------------------------------------------------
%

      We have been unable to find any relevant data (such as observations by existing spectroscopic surveys) on most of the stars in 
      Tables~\ref{best1}, \ref{vbest1}, and \ref{vbesth1}, but future spectroscopic studies of some of the candidates found here may be able 
      to confirm (or refute for false positives) an origin other than the Milky Way galaxy as discussed by \citet{2018MNRAS.481.1028H}. 
      The two exceptions are {\it Gaia}~DR2 2079869356551980672 and 1360405705322797312. {\it Gaia}~DR2 1360405705322797312 is a {\it bona 
      fide} member of the metal poor and very old globular cluster NGC~6341 or M~92 as its parameters match well those of Cl* NGC 6341 BARN 
      73 (for example, see \citealt{2000AJ....119.2895P,2014A&A...566A..58K}); therefore, it is a false positive (likely induced 
      by the crowded field) as the Galactocentric distance to NGC~6341 is 9.6~kpc, but the (wrong) value in Table~\ref{vbesth2} is 
      34.6$_{-14.9}^{+68.5}$~kpc. 
 
      In sharp contrast, the other previously studied object (meaning spectroscopically), {\it Gaia}~DR2 2079869356551980672 or KIC~8828284, 
      is given a spectral type of K5~II (bright giant) by \citet{2016A&A...594A..39F} and K2~Ib (supergiant) by 
      \citet{2016AJ....151...13G} with $V=14.23$~mag, which opens the door to a distance range from about 18~kpc (if bright giant with 
      $M_V=-2.0$~mag) to perhaps 60~kpc (if supergiant with $M_V=-4.7$~mag). Our data processing pipeline (see Table~\ref{vbest2}) gives a 
      value of the Galactocentric distance of 38.2$_{-14.2}^{+54.4}$~kpc for this object, which is consistent with the spectroscopic range, 
      and a Galactocentric velocity of 869$_{-342}^{+1269}$~km~s$^{-1}$ for an angle of 99\fdg8$_{-6\fdg1}^{+6\fdg0}$. It also gives 
      ${\rm G}_{\rm BP}-{\rm G}_{\rm RP}=1.5_{-0.7}^{+0.7}$ and ${\rm M}_{\rm G}=-5.1_{-2.4}^{+1.8}$~mag. In other words, if {\it Gaia}~DR2 
      2079869356551980672 is a giant, it might still be bound to the Galaxy but if it is a supergiant, it is a robust example of a passing 
      hypervelocity star with a velocity well above 1000~km~s$^{-1}$ and an origin outside the Milky Way galaxy.

      Regarding the other stars in Tables~\ref{best1}, \ref{vbest1}, and \ref{vbesth1}, it may be argued that we have not discussed the
      reliability of these candidates in terms of the Re-normalised Unit Weight Error (RUWE) criterion, which is based on the chi squared of
      the astrometric fit and stellar colour (see technical note GAIA-C3-TN-LU-LL-124-01),\footnote{\url{http://www.rssd.esa.int/doc_fetch.php?id=3757412}} 
      and can be found from a number of {\it Gaia}~DR2-related data servers.\footnote{\url{http://gaia.ari.uni-heidelberg.de/singlesource.html}} 
      \citet{2018A&A...616A...2L} argues that RUWE$\leq$1.4 is a reasonable criterion to identify sufficiently good solutions. The RUWE 
      index is included in Tables~\ref{best2}, \ref{vbest2}, and \ref{vbesth2} and all the sources have RUWE$<$1.4. Therefore, we confirm 
      that our best candidates have well-behaved astrometric solutions.

   \section{Conclusions\label{Concluding}}
      In this paper, we have investigated the spatial distribution of distant (nominal Galactocentric distances greater than 30~kpc)
      hypervelocity star candidates present in {\it Gaia} DR2 data in an attempt to understand the origin of the putative anisotropic 
      spatial distribution discussed by \citet{2009ApJ...690L..69B} for example, and confirm or refute the presence of a significant 
      population of high-velocity stars of intergalactic provenance as pointed out by \citet{2018MNRAS.tmp.2466M}. No global parallax 
      zero-point correction was performed. Our conclusions are:
      \begin{enumerate}[(i)]
         \item The effect of the Eddington-Trumpler-Weaver bias is very important for samples such as the ones studied here. We estimate 
               that over 80\% of the sources discussed in our work may be located considerably further away than implied by the values
               of their parallaxes and, therefore, most of our velocity estimates are just lower limits. 
         \item Most distant sources in {\it Gaia} DR2 have kinematics consistent with that of hypervelocity stars, but it cannot be 
               discarded that some candidates may actually belong to the foreground stellar populations (like in the case of the false 
               positive {\it Gaia}~DR2 1360405705322797312, which is a member of the globular cluster NGC~6341).
         \item We have identified hundreds of distant hypervelocity star candidates that might be experiencing fly-bys to the Milky Way 
               (their $V$-components are negative). The immediate neighbourhood of the Galaxy appears to host a significant population of
               exo-Galactic origin as pointed out by \citet{2018MNRAS.tmp.2466M} and such a population is consistent with findings by 
               \citet{2016A&A...588A..41C} for example, for a large sample of galaxies in the local Universe.
         \item We have been unable to identify a single hypervelocity star candidate with a probable origin in the region of the Galactic
               centre. The absence of such candidates is consistent with the results obtained by \citet{2018A&A...620A..48I} and 
               \citet{2018MNRAS.tmp.2466M} for other samples of {\it Gaia} DR2 sources with shorter Galactocentric distances.
         \item We have identified hundreds of distant high-velocity stars with a probable origin in the Galactic disc, confirming previous
               results by \citet{2018MNRAS.479.2789B}, \citet{2018ApJ...866..121H}, \citet{2018A&A...620A..48I}, 
               \citet{2018AJ....156...87L}, \citet{2019ApJ...873..116H}, and \citet{2018MNRAS.tmp.2466M}.
         \item We confirm that the spatial distribution of hypervelocity star candidates is anisotropic, but the origin of such an 
               anisotropy is probably the result of observational biases and selection effects.
         \item We identify a sample of candidate young, high-velocity stars with relatively low uncertainties that may have an intergalactic 
               provenance. One of them, {\it Gaia}~DR2 2079869356551980672 or KIC~8828284, emerges as a probable hypervelocity, supergiant 
               star with a velocity well above 1000~km~s$^{-1}$ and an origin outside the Milky Way galaxy. 
         \item Most distant hypervelocity star candidates identified here have probable ages under 50~Myr. This young age together with the
               values of their velocities argue for birth places located within 100~kpc from the centre of the Milky Way. 
      \end{enumerate}
      As a closing argument, it is probably worth to consider that in this work we have put the stress on the consequences of the 
      Eddington-Trumpler-Weaver bias, which leads to an underestimate of the heliocentric distance; however, we have to concede that the 
      existence of a negative zero-point offset in the {\it Gaia} DR2 parallax has the opposite effect, leading to an overestimate of the 
      heliocentric distance and, subsequently, the Galactocentric distance and velocities components. Which one, if any, is dominant in our 
      case? We may not be able to find an answer to this difficult question in absolute terms (not even with the future {\it Gaia} DR3). It 
      could be the case that the dominant effect (systematic overestimate due to {\it Gaia} parallax zero-point offset or underestimate 
      due to Eddington-Trumpler-Weaver bias) may depend on the region of the sky under study or both effects may cancel each other out 
      under certain circumstances or perhaps they follow some type of yet-to-be-found correlation. What appears to be clear is that the 
      presence in our sample of reasonably robust hypervelocity star candidates such as {\it Gaia}~DR2 2079869356551980672 (KIC~8828284) 
      argues against concluding that {\it Gaia} DR2 is not of sufficient quality to fulfill our original science goals: performing a 
      systematic exploration aimed at confirming or refuting that hypervelocity stars are ubiquitous in the local Universe and that, in the 
      Milky Way, the known hypervelocity stars are anisotropically distributed. 
      
   \begin{acknowledgements}
      We thank the referee for her/his constructive, actionable, and insightful reports that included very helpful suggestions regarding the 
      presentation of this paper and the interpretation of our results, S.~J. Aarseth for comments on black-hole-induced ejections and A.~I. 
      G\'omez de Castro for comments on high-velocity stars and for providing access to computing facilities; RdlFM thanks L. Beitia-Antero 
      for extensive discussions on {\it Gaia} DR2 data and F.~J. Y\'a\~nez Gestoso for comments on some of the statistical issues discussed 
      in this work. This research was partially supported by the Spanish `Ministerio de Econom\'{\i}a y Competitividad' (MINECO) under 
      grants ESP2015-68908-R and ESP2017-87813-R. In preparation of this paper, we made use of the NASA Astrophysics Data System, the 
      ASTRO-PH e-print server, and the SIMBAD and VizieR databases operated at CDS, Strasbourg, France. This work has made use of data from 
      the European Space Agency (ESA) mission {\it Gaia} (\url{https://www.cosmos.esa.int/gaia}), processed by the {\it Gaia} Data 
      Processing and Analysis Consortium (DPAC, \url{https://www.cosmos.esa.int/web/gaia/dpac/consortium}). Funding for the DPAC has been 
      provided by national institutions, in particular the institutions participating in the {\it Gaia} Multilateral Agreement.
   \end{acknowledgements}

   \bibliographystyle{aa}

\begin{thebibliography}{}
      \bibitem[Aarseth(2012)]{2012MNRAS.422..841A} Aarseth, S.~J.\ 2012, 
              \mnras, 422, 841
      \bibitem[Abuter et al.(2019)]{2019arXiv190405721A} Abuter, R., Amorim, A., Bauboeck, M., et al.\ 2019, 
              \aap, submitted (arXiv:1904.05721)
      \bibitem[Andrae et al.(2018)]{2018A&A...616A...8A} Andrae, R., Fouesneau, M., Creevey, O., et al.\ 2018,
              \aap, 616, A8
      \bibitem[Arenou et al.(2018)]{2018A&A...616A..17A} Arenou, F., Luri, X., Babusiaux, C., et al.\ 2018, 
              \aap, 616, A17
      \bibitem[Astraatmadja \& Bailer-Jones(2016)]{2016ApJ...832..137A} Astraatmadja, T.~L., \& Bailer-Jones, C.~A.~L.\ 2016, 
              \apj, 832, 137
      \bibitem[Bailer-Jones(2015)]{2015PASP..127..994B} Bailer-Jones, C.~A.~L.\ 2015, 
              \pasp, 127, 994
      \bibitem[Bailer-Jones et al.(2018)]{2018AJ....156...58B} Bailer-Jones, C.~A.~L., Rybizki, J., Fouesneau, M., Mantelet, G., \& Andrae, R.\ 2018, 
              \aj, 156, 58
      \bibitem[Bell et al.(1991)]{1991MNRAS.250..119B} Bell, S.~A., Hill, G., Hilditch, R.~W., et al.\ 1991, 
              \mnras, 250, 119
      \bibitem[Binney \& Merrifield(1998)]{1998gaas.book.....B} Binney, J., \& Merrifield, M.\ 1998, 
              Galactic Astronomy, 
              Princeton, NJ: Princeton University Press
      \bibitem[Bonning et al.(2007)]{2007ApJ...666L..13B} Bonning, E.~W., Shields, G.~A., \& Salviander, S.\ 2007, 
              \apjl, 666, L13
      \bibitem[Boubert et al.(2017)]{2017MNRAS.469.2151B} Boubert, D., Erkal, D., Evans, N.~W., \& Izzard, R.~G.\ 2017, 
              \mnras, 469, 2151
      \bibitem[Boubert et al.(2018)]{2018MNRAS.479.2789B} Boubert, D., Guillochon, J., Hawkins, K., et al.\ 2018, 
              \mnras, 479, 2789
      \bibitem[Boubert et al.(2019)]{2019MNRAS.486.2618B} Boubert, D., Strader, J., Aguado, D., et al.\ 2019, 
              \mnras, 486, 2618
      \bibitem[Box \& Muller(1958)]{BM58} Box, G.~E.~P., \& Muller, M.~E.\ 1958,
              Ann. Math. Stat., 29, 610
      \bibitem[Bressan et al.(2012)]{2012MNRAS.427..127B} Bressan, A., Marigo, P., Girardi, L., et al.\ 2012, 
              \mnras, 427, 127
      \bibitem[Bromley et al.(2018)]{2018ApJ...868...25B} Bromley, B.~C., Kenyon, S.~J., Brown, W.~R., \& Geller, M.~J.\ 2018, 
              \apj, 868, 25
      \bibitem[Brown(2015)]{2015ARA&A..53...15B} Brown, W.~R.\ 2015, 
              \araa, 53, 15
      \bibitem[Brown et al.(2005)]{2005ApJ...622L..33B} Brown, W.~R., Geller, M.~J., Kenyon, S.~J., \& Kurtz, M.~J.\ 2005, 
              \apjl, 622, L33
      \bibitem[Brown et al.(2009)]{2009ApJ...690L..69B} Brown, W.~R., Geller, M.~J., Kenyon, S.~J., \& Bromley, B.~C.\ 2009, 
              \apjl, 690, L69
      \bibitem[Caldwell et al.(2014)]{2014ApJ...787L..11C} Caldwell, N., Strader, J., Romanowsky, A.~J., et al.\ 2014, 
              \apjl, 787, L11
      \bibitem[Cicone et al.(2016)]{2016A&A...588A..41C} Cicone, C., Maiolino, R., \& Marconi, A.\ 2016, 
              \aap, 588, A41
      \bibitem[Dame \& Thaddeus(2011)]{2011ApJ...734L..24D} Dame, T.~M., \& Thaddeus, P.\ 2011, 
              \apjl, 734, L24
      \bibitem[de la Fuente Marcos \& de la Fuente Marcos(2008a)]{2008ApJ...677L..47D} de la Fuente Marcos, R., \& de la Fuente Marcos, C.\ 2008a, 
              \apjl, 677, L47
      \bibitem[de la Fuente Marcos \& de la Fuente Marcos(2008b)]{2008ApJ...685L.125D} de la Fuente Marcos, R., \& de la Fuente Marcos, C.\ 2008b, 
              \apjl, 685, L125
      \bibitem[de la Fuente Marcos \& de la Fuente Marcos(2018)]{2018MNRAS.481L..64D} de la Fuente Marcos, R., \& de la Fuente Marcos, C.\ 2018, 
              \mnras, 481, L64
      \bibitem[Deason et al.(2019)]{2019MNRAS.485.3514D} Deason, A.~J., Fattahi, A., Belokurov, V., et al.\ 2019, 
              \mnras, 485, 3514
      \bibitem[Du et al.(2018)]{2018ApJ...869L..31D} Du, C., Li, H., Newberg, H.~J., et al.\ 2018, 
              \apjl, 869, L31 
      \bibitem[Eddington(1913)]{1913MNRAS..73..359E} Eddington, A.~S.\ 1913, 
              \mnras, 73, 359
      \bibitem[Eddington(1940)]{1940MNRAS.100..354E} Eddington, A.~S.\ 1940, 
              \mnras, 100, 354
      \bibitem[Erkal et al.(2019)]{2019MNRAS.483.2007E} Erkal, D., Boubert, D., Gualandris, A., Evans, N.~W., \& Antonini, F.\ 2019, 
              \mnras, 483, 2007
      \bibitem[Francis(2014)]{2014MNRAS.444L...6F} Francis, C.\ 2014, 
              \mnras, 444, L6
      \bibitem[Frasca et al.(2016)]{2016A&A...594A..39F} Frasca, A., Molenda-{\.Z}akowicz, J., De Cat, P., et al.\ 2016, 
              \aap, 594, A39
      \bibitem[Freedman \& Diaconis(1981)]{FD81} Freedman, D., \& Diaconis, P.\ 1981,
              Z. Wahrscheinlichkeitstheor. verwandte Geb., 57, 453
      \bibitem[Gaia Collaboration, Prusti et al.(2016)]{2016A&A...595A...1G} Gaia Collaboration, Prusti, T., de Bruijne, J.~H.~J., et al.\ 2016, 
              \aap, 595, A1
      \bibitem[Gaia Collaboration, Brown et al.(2018)]{2018A&A...616A...1G} Gaia Collaboration, Brown, A.~G.~A., Vallenari, A., et al.\ 2018,
              \aap, 616, A1
      \bibitem[Gaia Collaboration, Babusiaux et al.(2018)]{2018A&A...616A..10G} Gaia Collaboration, Babusiaux, C., van Leeuwen, F., et al.\ 2018,
              \aap, 616, A10
      \bibitem[Gaia Collaboration, Helmi et al.(2018)]{2018A&A...616A..12G} Gaia Collaboration, Helmi, A., van Leeuwen, F., et al.\ 2018, 
              \aap, 616, A12
      \bibitem[Gal-Yam et al.(2003)]{2003AJ....125.1087G} Gal-Yam, A., Maoz, D., Guhathakurta, P., \& Filippenko, A.~V.\ 2003, 
              \aj, 125, 1087
      \bibitem[Graczyk et al.(2019)]{2019ApJ...872...85G} Graczyk, D., Pietrzy{\'n}ski, G., Gieren, W., et al.\ 2019, 
              \apj, 872, 85
      \bibitem[Gray et al.(2016)]{2016AJ....151...13G} Gray, R.~O., Corbally, C.~J., De Cat, P., et al.\ 2016, 
              \aj, 151, 13
      \bibitem[Gualandris \& Merritt(2008)]{2008ApJ...678..780G} Gualandris, A., \& Merritt, D.\ 2008, 
              \apj, 678, 780
      \bibitem[Guillochon et al.(2017)]{2017ApJ...835...64G} Guillochon, J., Parrent, J., Kelley, L.~Z., \& Margutti, R.\ 2017, 
              \apj, 835, 64
      \bibitem[Hattori et al.(2018)]{2018ApJ...866..121H} Hattori, K., Valluri, M., Bell, E.~F., \& Roederer, I.~U.\ 2018, 
              \apj, 866, 121
      \bibitem[Hattori et al.(2019)]{2019ApJ...873..116H} Hattori, K., Valluri, M., Castro, N., et al.\ 2019, 
              \apj, 873, 116
      \bibitem[Hawkins \& Wyse(2018)]{2018MNRAS.481.1028H} Hawkins, K., \& Wyse, R.~F.~G.\ 2018, 
              \mnras, 481, 1028
      \bibitem[Hilditch et al.(2005)]{2005MNRAS.357..304H} Hilditch, R.~W., Howarth, I.~D., \& Harries, T.~J.\ 2005, 
              \mnras, 357, 304
      \bibitem[Hills(1988)]{1988Natur.331..687H} Hills, J.~G.\ 1988, 
              \nat, 331, 687
      \bibitem[Huang et al.(2017)]{2017ApJ...847L...9H} Huang, Y., Liu, X.-W., Zhang, H.-W., et al.\ 2017, 
              \apjl, 847, L9
      \bibitem[Ibata et al.(1994)]{1994Natur.370..194I} Ibata, R.~A., Gilmore, G., \& Irwin, M.~J.\ 1994, 
              \nat, 370, 194
      \bibitem[Irrgang et al.(2018)]{2018A&A...620A..48I} Irrgang, A., Kreuzer, S., \& Heber, U.\ 2018, 
              \aap, 620, A48
      \bibitem[Johnson \& Soderblom(1987)]{1987AJ.....93..864J} Johnson, D.~R.~H., \& Soderblom, D.~R.\ 1987, 
              \aj, 93, 864
      \bibitem[Kamdar et al.(2019)]{2019arXiv190402159K} Kamdar, H., Conroy, C., Ting, Y.-S., et al.\ 2019, 
              ApJL, submitted (arXiv:1904.02159)
      \bibitem[Kamann et al.(2014)]{2014A&A...566A..58K} Kamann, S., Wisotzki, L., Roth, M.~M., et al.\ 2014, 
              \aap, 566, A58
      \bibitem[Koen(1992)]{1992MNRAS.256...65K} Koen, C.\ 1992, 
              \mnras, 256, 65
      \bibitem[Kogut et al.(1993)]{1993ApJ...419....1K} Kogut, A., Lineweaver, C., Smoot, G.~F., et al.\ 1993, 
              \apj, 419, 1
      \bibitem[Leung \& Bovy(2019)]{2019arXiv190208634L} Leung, H.~W., \& Bovy, J.\ 2019, 
              \mnras, submitted (arXiv:1902.08634)
      \bibitem[Li et al.(2018)]{2018AJ....156...87L} Li, Y.-B., Luo, A.-L., Zhao, G., et al.\ 2018, 
              \aj, 156, 87
      \bibitem[Lindegren et al.(2018)]{2018A&A...616A...2L} Lindegren, L., Hern{\'a}ndez, J., Bombrun, A., et al.\ 2018, 
              \aap, 616, A2
      \bibitem[Luri et al.(2018)]{2018A&A...616A...9L} Luri, X., Brown, A.~G.~A., Sarro, L.~M., et al.\ 2018, 
              \aap, 616, A9
      \bibitem[Lutz \& Kelker(1973)]{1973PASP...85..573L} Lutz, T.~E., \& Kelker, D.~H.\ 1973, 
              \pasp, 85, 573
      \bibitem[Marchetti et al.(2019)]{2018MNRAS.tmp.2466M} Marchetti, T., Rossi, E.~M., \& Brown, A.~G.~A.\ 2019, 
              \mnras, in press (10.1093/mnras/sty2592)
      \bibitem[Marigo et al.(2017)]{2017ApJ...835...77M} Marigo, P., Girardi, L., Bressan, A., et al.\ 2017,
              \apj, 835, 77
      \bibitem[Metropolis \& Ulam(1949)]{MU49} Metropolis, N., \& Ulam, S.\ 1949,
              J. Am. Stat. Assoc., 44, 335
      \bibitem[Mikkola \& Valtonen(1990)]{1990ApJ...348..412M} Mikkola, S., \& Valtonen, M.~J.\ 1990, 
              \apj, 348, 412
      \bibitem[Muraveva et al.(2014)]{2014MNRAS.443..432M} Muraveva, T., Clementini, G., Maceroni, C., et al.\ 2014, 
              \mnras, 443, 432
      \bibitem[Ochsenbein et al.(2000)]{2000A&AS..143...23O} Ochsenbein, F., Bauer, P., \& Marcout, J.\ 2000, 
              \aaps, 143, 23
      \bibitem[Oudmaijer et al.(1998)]{1998MNRAS.294L..41O} Oudmaijer, R.~D., Groenewegen, M.~A.~T., \& Schrijver, H.\ 1998, 
              \mnras, 294, L41
      \bibitem[Pastorelli et al.(2019)]{2019MNRAS.485.5666P} Pastorelli, G., Marigo, P., Girardi, L., et al.\ 2019, 
              \mnras, 485, 5666
      \bibitem[Pawlak et al.(2013)]{2013AcA....63..323P} Pawlak, M., Graczyk, D., Soszy{\'n}ski, I., et al.\ 2013, 
              \actaa, 63, 323 
      \bibitem[Pietrzy{\'n}ski et al.(2013)]{2013Natur.495...76P} Pietrzy{\'n}ski, G., Graczyk, D., Gieren, W., et al.\ 2013, 
              \nat, 495, 76
      \bibitem[Pilachowski et al.(2000)]{2000AJ....119.2895P} Pilachowski, C.~A., Sneden, C., Kraft, R.~P., Harmer, D., \& Willmarth, D.\ 2000, 
              \aj, 119, 2895
      \bibitem[Press et al.(2007)]{2002nrca.book.....P} Press, W.~H., Teukolsky, S.~A., Vetterling, W.~T., \& Flannery, B.~P.\ 2007, 
              Numerical Recipes: The Art of Scientific Computing, 3rd edn., 
              Cambridge, UK: Cambridge University Press  
      \bibitem[Reid et al.(2014)]{2014ApJ...783..130R} Reid, M.~J., Menten, K.~M., Brunthaler, A., et al.\ 2014, 
              \apj, 783, 130
      \bibitem[Riess et al.(2018)]{2018ApJ...861..126R} Riess, A.~G., Casertano, S., Yuan, W., et al.\ 2018, 
              \apj, 861, 126
      \bibitem[Saslaw \& De Young(1972)]{1972ApL....11...87S} Saslaw, W.~C., \& De Young, D.~S.\ 1972, 
              \aplett, 11, 87
      \bibitem[Saslaw et al.(1974)]{1974ApJ...190..253S} Saslaw, W.~C., Valtonen, M.~J., \& Aarseth, S.~J.\ 1974, 
              \apj, 190, 253
      \bibitem[Schlafly \& Finkbeiner(2011)]{2011ApJ...737..103S} Schlafly, E.~F., \& Finkbeiner, D.~P.\ 2011, 
              \apj, 737, 103
      \bibitem[Schlegel et al.(1998)]{1998ApJ...500..525S} Schlegel, D.~J., Finkbeiner, D.~P., \& Davis, M.\ 1998, 
              \apj, 500, 525
      \bibitem[Sch{\"o}nrich et al.(2010)]{2010MNRAS.403.1829S} Sch{\"o}nrich, R., Binney, J., \& Dehnen, W.\ 2010, 
              \mnras, 403, 1829
      \bibitem[Sch{\"o}nrich et al.(2011)]{2011MNRAS.415.3807S} Sch{\"o}nrich, R., Asplund, M., \& Casagrande, L.\ 2011, 
              \mnras, 415, 3807
      \bibitem[Sch{\"o}nrich et al.(2019)]{2019MNRAS.tmp.1390S} Sch{\"o}nrich, R., McMillan, P., \& Eyer, L.\ 2019, 
              \mnras, in press (10.1093/mnras/stz1451)
      \bibitem[Sharma et al.(2011)]{2011ApJ...730....3S} Sharma, S., Bland-Hawthorn, J., Johnston, K.~V., \& Binney, J.\ 2011, 
              \apj, 730, 3
      \bibitem[Shen et al.(2018)]{2018ApJ...865...15S} Shen, K.~J., Boubert, D., G{\"a}nsicke, B.~T., et al.\ 2018, 
              \apj, 865, 15      
      \bibitem[Smith(2003)]{2003MNRAS.338..891S} Smith, H.\ 2003, 
              \mnras, 338, 891
      \bibitem[Stassun \& Torres(2018)]{2018ApJ...862...61S} Stassun, K.~G., \& Torres, G.\ 2018, 
              \apj, 862, 61
      \bibitem[Tauris(2015)]{2015MNRAS.448L...6T} Tauris, T.~M.\ 2015, 
              \mnras, 448, L6
      \bibitem[Trumpler \& Weaver(1953)]{1953stas.book.....T} Trumpler, R.~J., \& Weaver, H.~F.\ 1953, 
              Statistical Astronomy,
              Dover Books on Astronomy and Space Topics, New York: Dover Publications
      \bibitem[Wall \& Jenkins(2012)]{2012psa..book.....W} Wall, J.~V., \& Jenkins, C.~R.\ 2012,
              Practical Statistics for Astronomers,
              Cambridge, UK: Cambridge University Press
      \bibitem[Xu et al.(2019)]{2019ApJ...875..114X} Xu, S., Zhang, B., Reid, M.~J., Zheng, X., \& Wang, G.\ 2019, 
              \apj, 875, 114
      \bibitem[Yu \& Tremaine(2003)]{2003ApJ...599.1129Y} Yu, Q., \& Tremaine, S.\ 2003, 
              \apj, 599, 1129
      \bibitem[Zinn et al.(2019)]{2018arXiv180502650Z} Zinn, J.~C., Pinsonneault, M.~H., Huber, D., \& Stello, D.\ 2019, 
              \apj, in press (arXiv:1805.02650)
   \end{thebibliography}

   \begin{appendix}
      \section{Effects of a systematic parallax zero-point offset\label{ZPoffset}}
         As discussed in Sect.~\ref{DR2ZP}, there is admittedly robust evidence for the existence of a systematic negative parallax 
         zero-point offset in {\it Gaia} DR2. This leads to values of the parallaxes that are too small, even after Monte Carlo sampling. 
         Here, we used its most optimistic determination, $-0.029$~mas \citep{2018A&A...616A...2L} that is based on quasars and was
         criticised by \citet{2019MNRAS.tmp.1390S} among others, to study its effects on our conclusions. In other words, we repeated
         the calculations described above but using $d_{\rm c}=1/(\pi_{\rm c}+0.029~{\rm mas})$ instead of the usual $d_{\rm c}=1/\pi_{\rm c}$. 
         The immediate effect of this choice is that not a single one of the 174 high-velocity candidates (see Sect.~\ref{BestRes}) with the 
         most reliable astrometric solutions survives as such after performing the parallax zero-point offset correction. However, 121$\pm$2 
         of the 15681 sources (nearly 0.8\%) in our primary sample are still hypervelocity star candidates according to our criteria ---at a 
         Galactocentric distance $>30$~kpc, the Galactocentric velocity must be $>500$~km~s$^{-1}$--- after performing the correction, but 
         these are all sources with somewhat questionable astrometric solutions as they pass some quality-control criteria but not all of 
         them (see Sect.~\ref{SeCr}). In sharp contrast and when no correction was applied, the fraction of hypervelocity star candidates is 
         about 30\% (4774$\pm$8 out of 15681). In other words, 97\% of the hypervelocity star candidates cannot be considered as such when 
         the correction is performed. This fraction is comparable to the strength of the effect of the Eddington-Trumpler-Weaver bias 
         discussed in Sect.~\ref{FarAway}, over 80\% of the sources studied in our work may be located further away than implied by the 
         values of their parallaxes.

         If instead of using $-0.029$~mas, we consider the most pessimistic determination, $-0.082\pm0.033$~mas \citep{2018ApJ...862...61S}, 
         then the number of high-velocity candidates goes down to zero. The same result is obtained if we apply the zero-point offset 
         computed by \citet{2019MNRAS.tmp.1390S}, -0.054$\pm$0.006~mas. This outcome was expected because the 87\,733\,672 sources with
         estimated values of the line-of-sight extinction and reddening in {\it Gaia} DR2 have strictly positive values of the parallax and 
         these values of the zero-point offset (in absolute terms) are above the cutoff value 0.033~mas (or 30~kpc). This fact also explains 
         why a small fraction of high-velocity candidates manage to survive the correction process when the most optimistic determination, 
         $-0.029$~mas, is used.

         Therefore, if one believes that not having carried out the zero-point offset correction is a serious weakness in our analysis, then 
         one must conclude that {\it Gaia} DR2 may not be not adequate to fulfill our original science goals, perhaps the future {\it Gaia} 
         DR3 will be better suited for the task, but see our closing argument in Sect.~\ref{Concluding}.
   \end{appendix}

\end{document}